\documentclass[fleqn,usenatbib]{mnras}

\usepackage{newtxtext,newtxmath}

\usepackage[LGRgreek]{mathastext} 

\usepackage[T1]{fontenc}

\DeclareRobustCommand{\VAN}[3]{#2}
\let\VANthebibliography\thebibliography
\def\thebibliography{\DeclareRobustCommand{\VAN}[3]{##3}\VANthebibliography}

\usepackage{graphicx}	
\usepackage{amsmath}	

\usepackage{amssymb}	
\usepackage{longtable}
\usepackage{caption}
\usepackage{subcaption}
\usepackage{pdflscape}	

\title[Feedback in MaNGA AGN]{The extent and power of "maintainance mode" feedback in MaNGA AGN}

\author[Lara Gatto]{Lara Gatto,$^{1}$\thanks{E-mail:lara.gatto@ufrgs.br}
T. Storchi-Bergmann,$^{1,2}$ Rogemar A. Riffel,$^{2,3}$ Rogério Riffel,$^{1,2,4}$ Sandro B. Rembold,$^{2,3}$ \newauthor Jaderson S. Schimoia,$^{2,3}$ Nicolas D. Mallmann,$^{1,2}$ Gabriele S. Ilha,$^{5,6}$
\\
$^{1}$ Instituto de Física, Universidade Federal do Rio Grande do Sul, Av. Bento Gonçalves 9500, 91501-970 Porto Alegre, RS, Brazil\\
$^{2}$ Laboratório Interinstitucional de e-Astronomia - LIneA, Rua Gal. José Cristino 77, Rio de Janeiro, RJ - 20921-400, Brazil\\
$^{3}$ Departamento de F\'\i sica, Centro de Ci\^encias Naturais e Exatas, Universidade Federal de Santa Maria, 97105-900, Santa Maria, RS, Brazil \\ 
$^{4}$ Instituto de Astrof\'\i sica de Canarias, Calle V\'\i a L\'actea s/n, E-38205 La Laguna, Tenerife, Spain\\
$^{5}$ Universidade de São Paulo, Instituto de Astronomia,
Geofísica e Ciências Atmosféricas, Rua do Matão 1226, CEP 05508-090, São Paulo, SP, Brazil\\
$^{6}$ Universidade do Vale do Para\'iba. Av. Shishima Hifumi, 2911, CEP: 12244-000, São Jos\'e dos Campos, SP, Brazil\\
}

\date{Accepted XXX. Received YYY; in original form ZZZ}

\pubyear{2023}

\begin{document}
\label{firstpage}
\pagerange{\pageref{firstpage}--\pageref{lastpage}}
\maketitle

\begin{abstract}
We study the ionised gas kinematics of 293 Active Galactic Nuclei (AGN) hosts as compared to that of 485 control galaxies from the MaNGA-SDSS survey using measurements of the [O\,{\sc iii}]$\lambda$5007\AA\, emission-line profiles, presenting flux, velocity and W$_{80}$ maps. In 45\%\ of the AGN, a broad component was needed to fit the line profiles wings within the inner few kpc, that we have identified with an outflow. But in most AGN, the profiles are broader than that of their controls over a much more extended region, identified as the "kinematically disturbed regions" (KDRs). We find a positive correlation between the mean $\langle$W$_{80}\rangle$ and L[O\,{\sc III}], supporting that the KDR is due to heating and turbulence of the ISM by outflows and radiation from the AGN. The extent R$_{KDR}$ reaches up to 24\,kpc, with a mean ratio to that of the ENLR of 57\%. We estimate ionised gas mass flow rates ($\dot{M}_{\rm out}$) and kinetic powers ($\dot{E}_{\rm out}$) both from the AGN broad components and from the W$_{80}$ values, that can be obtained for the whole AGN sample. We find values for $\dot{M}_{\rm out}$ and $\dot{E}_{\rm out}$ that correlate with the AGN luminosity $L_{bol}$, populating the low luminosity end of these known correlations. The mean coupling efficiency between $\dot{E}_{\rm out}$ and AGN luminosity is $\approx$ 0.02\% from the W$_{80}$ values and lower from the broad component.
But the large extent of the KDR shows that even low-luminosity AGN can impact the host galaxy along several kpc in a "maintenance mode" feedback.
\end{abstract}




\section{Introduction}

The vast majority of galaxies that present a stellar bulge host a supermassive black hole (SMBHs) in their centers. If the SMBH is capturing mass from its surroundings \citep{Thaisa_2019}, feeding the SMBH via an accretion disk, the nucleus of the galaxy becomes an Active Galactic Nucleus (AGN) \citep{Antonucci,kormendy_HO_13}. 

The accretion process produces energy via radiation, jets and outflows from the accretion disk. If this energy is efficiently coupled with the gas present in the central region and sometimes up to galactic scales \citep{Costa_20}, the AGN can have a significant impact -- referred to as feedback, on their host galaxies. Negative feedback from the AGN can heat the star forming material and/or eventually expel the gas resulting in a shut down of star formation \citep{Choi_2018}. These processes end up regulating the galaxy growth and avoiding the build-up of over-massive galaxies \citep{Fabian_2012,McNamara_2012,nelson_19}. Eventually, the AGN can cause a positive feedback too, by triggering star formation and impacting the surrounding intergalactic medium \citep[e.g.][]{maiolino17,gallagher}.

In powerful AGN, the main observable interaction with the surrounding gas is the presence of kinetic disturbances such as outflows or velocity dispersion enhancements, both along and perpendicularly to radio jets and ionisation axes \citep[e.g.][]{Rogemar_2015,Venturi_2021,2021_Rifel_chemicalAbundance}. The kinetic feedback on the gas surrounding AGN comprise complex and multi-phase phenomena, with different gas phases observable in different spectral domains \citep{Roy_2021}. Most of our current knowledge about AGN-driven kinetic feedback comes from mapping the kinematics of the warm ionised gas phase via optical emission lines such as [O\,{\sc iii}]$\lambda$ 5007\AA\, \citep[e.g.][]{Dominika,Bruno_21,kakkad_2022}. 

Integral field unit (IFU) surveys of galaxies have been allowing the characterisation of kinetic signatures such as outflows and velocity dispersion enhancements for statistically significant samples, such as the SDSS-IV survey MaNGA -- Mapping Nearby Galaxies at Apache Point Observatory (APO), an optical fiber-bundle IFU survey that has obtained data of 10,000 galaxies at z $\leq$ 0.1 \citep{bundy,Dr_17_SDSS}. Another wide-field IFU survey of galaxies that can be mentioned is the Calar Alto Legacy Integral Field Area Survey (CALIFA) that has observed a sample of ~600 galaxies in the local universe \citep{Califa}.

The present work is part of a series of studies of our group AGNIFS - AGN Integral Field Spectroscopy on MaNGA AGN to investigate the effect of AGN on their host galaxies. With this goal, we have defined the AGN and control samples according to criteria described In Paper I \citep{Sandro-PAPER1}, applied also in subsequent studies of the series. Paper II \citep{Mallmann_paperII} presented spatially resolved stellar population properties, showing that the fraction of young stars in high-luminosity AGN is higher in the inner regions when compared with the control sample. Paper III \citep{Jana_paperIV} focused on the emission-line flux distributions and gas excitation and reported that the extent of the region ionised by the AGN is proportional to $L^{0.5}_{\rm [O\:III]}$, where  $L_{\rm [O\:III]}$ is the nuclear luminosity of the  [O\,{\sc iii}]$\lambda$5007\AA\, emission line. They also found that the star formation rate (SFR) is higher in AGN than in the control galaxies for the early-type host galaxies. The analysis of the nuclear stellar and gas kinematics, as well as the difference of the orientation of the line of nodes derived from the stellar and gas velocity fields (kinematic position angle [PA] offset), was presented in Paper IV \citep{ilha_paperIII}. By comparing AGN host and inactive galaxies, no difference was found in terms of the kinematic PA offsets between gas and stars. But AGN present higher gas velocity dispersion within the inner 2$\farcs$5 diameter region, that has been interpreted as due to AGN-driven kinetic disturbances/outflows. 
Paper V \citep{Alice_paperV} characterised and estimated the extents of the Narrow Line Region (NLR) and of the kinematically disturbed region (KDR) by the AGN in a sample of 173 AGN host galaxies and corresponding controls, adopting the gas velocity dispersion (the $\sigma$ of the fitted Gaussian curve to the line) as an indicator of this disturbance. They found that the extension of the KDR corresponds, on average, to about 30 per cent of that of the NLR. Assuming that the KDR is due to an AGN outflow, they estimated ionised gas mass outflow rates between 10$^{-5}$ and 1 M$_{\sun}$ yr$^{-1}$, and kinetic powers that range from 10$^{34}$ to 10$^{40}$\,erg\,s$^{-1}$. In \citet[][, Paper VI]{Rogerio_21_sfr}, the SFR was obtained via stellar population synthesis and found to be proportional to that obtained from the gas for HII regions in the sample galaxies, and a relation was obtained between the two, that can be used to obtain the SFR in regions ionised by the AGN.

In the present study, we investigate further the ionised gas kinematics, now of the complete sample, comprising 293 AGN and 485 control galaxies. Instead of using the gas velocity dispersion $\sigma$ as an indicator of the kinematic disturbance introduced by the AGN, as in \citet{Alice_paperV}, we use here the non-parametric measure W$_{80}$ \citep[e.g.][]{Whittle85,Veilleux_1991b,Dominika} of the  [O\,{\sc iii}]$\lambda$ 5007\AA\ emission line, that is more sensitive than $\sigma$ to the highest velocity contribution present in the profile wings. 
We also call the region in which the AGN produces kinematic disturbances in the gas as KDR, as in Paper V and obtain measurements of its extent, as well as of the extent of the Extended NLR (ENLR). Finally, we obtain the gas masses, mass flow rates and the kinetic feedback power of the AGN on the host galaxies.

This paper is organised as follows: Sec. \ref{sec_2} introduces our data. Sec. \ref{sec_3} presents the spectroscopic fitting procedures and measurements. Sec. \ref{sec_4} presents the results comprising flux and kinematic maps, including the distribution of W$_{80}$ values. In Sec. \ref{sec:5} we discus the W$_{80}$ maps and relation to other properties, we define the KDR, and estimate the AGN kinetic feedback power. In Sec.\ref{sec:6}  we present our conclusions. The assumed cosmological parameters in this work are H$_0 = 70$\,km\,s$^{-1}$\,Mpc$^{-1}$, $\Omega_m = 0.3$, and $\Omega_V = 0.7$.

\section{Data}
\label{sec_2}

The Mapping Nearby Galaxies at APO (MaNGA) survey \citep{bundy}, part of the fourth generation Sloan Digital Sky Survey (SDSS IV) is an Integral-Field Spectroscopic Survey that has obtained resolved optical (3600 \AA\,-10400 \AA) spectroscopy of $\sim$ 10,000 galaxies in the near Universe (with mean $z \approx $ 0.03). For details about the observations, see \cite{Drory_2015,Law_2015,Yan_2016_a,Yan_2016_b}. The data used in this present paper is from the MaNGA final data release (DR17, \citep{Dr_17_SDSS}). 

The sample of AGN galaxies was selected following the methodology described in \citet{Sandro-PAPER1}, using the BPT \citep{Baldwin_1981} and WHAN \citep{cid2010,Cid2011} diagnostic diagrams to select the AGN on the basis of the nuclear spectra. The WHAN diagram was used in order to
eliminate from the AGN sample the possible “LIERs”, or “fake AGN”, corresponding to galaxies that have H$\alpha$ equivalent widths EW$_{H\alpha}$ smaller than 3\AA\ which can be ionised by hot stars instead of AGN \citep{Cid2011}. The total number of AGN in MaNGA, according to our criteria, is initially, of 298 galaxies.
In order to characterise the properties of the host galaxy due to the presence of an AGN, we have chosen two control galaxies to match each AGN host, according to the galaxy morphology, stellar mass, redshift and galaxy inclination.
Since, in a few cases, we had to use the same control galaxy for more than one AGN host, this control sample is composed by 485 sources (not exactly twice the number of AGN).  Both AGN and control samples present redshifts in the range 0.013  $\leq$ z $\leq$ 0.15, and their typical stellar masses are in the range 10$^{10.5}$ -- 10$^{11}$ M$_{\odot}$. 

\begin{figure*}
    \centering
    \includegraphics[trim = 45mm 0mm 40mm 05mm,clip,width=\textwidth]{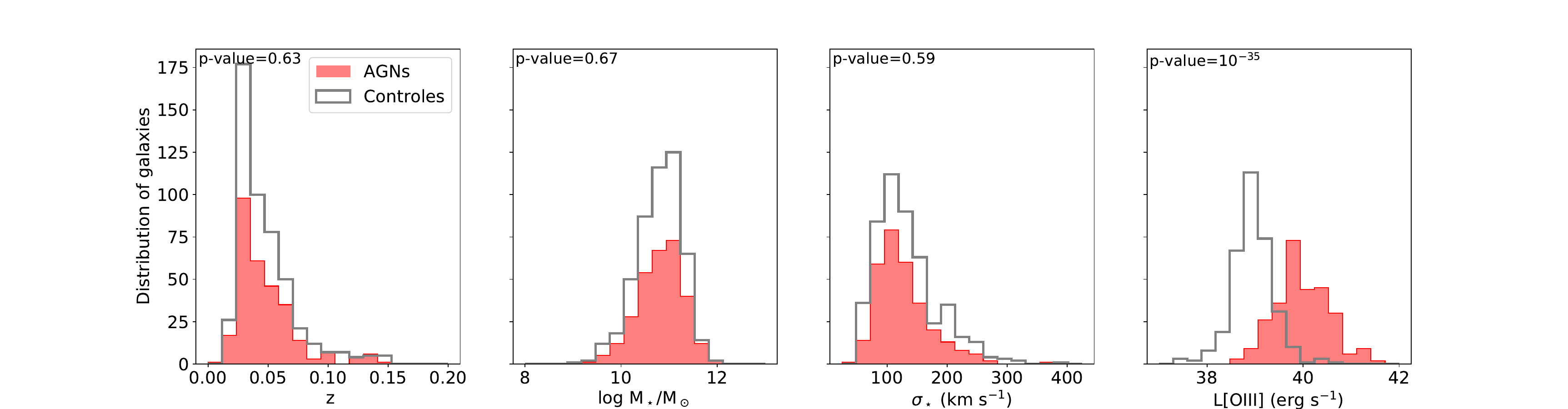}
    \caption{From left to right: distributions of redshift, stellar mass, stellar velocity dispersion, and [O\,{\sc iii}]$\lambda$5007\AA luminosity of the AGN host sample (filled red histogram) and control sample (empty gray histogram). The p-value shown refers to the KS-test. This figure is based on figure 1 of \citet{Riffel23_megacube}.}
    \label{fig:hist_sample}
\end{figure*}

In figure \ref{fig:hist_sample} we show the distribution of redshifts, stellar masses, stellar velocity dispersions, and [O\,{\sc iii}]$\lambda$5007\AA\, luminosities for the AGN and control sample, adapted from \citet{Riffel23_megacube}. The AGN and control samples present markedly different distributions only for the $L_{\rm [O\:III]}$, with the distributions of all the other parameters well matched between the two samples, as given by the low p-value of the KS-test, shown in the figure. We point out, in particular, the similar distributions for the stellar velocity dispersion, that assures that any kinematic difference found in the gas, between AGN and controls, is not due to differences in the host's galactic potential.

Considering the fact that the main goal of this paper is to compare the gas kinematics of AGN hosts and their control galaxies, for five AGN in our sample it was not possible to find suitable control galaxies, thus we do not include these objects in the global statistics. The final AGN sample thus comprises 293 galaxies. Most of these AGN (85\%) are low-luminosity, presenting [O\,{\sc iii}]$\lambda$5007\AA\ luminosities below $3.8 \times 10^{40}$\,ergs\,s$^{-1}$.

We have looked for the morphological classification of the galaxies in the Galaxy Zoo \citep{Lintott_galax_zoo,galaxy_zoo} database and found that our AGN hosts are comprised by 60\% spiral galaxies, 32\% elliptical and 5\% merging galaxies and the remaining 3\% are not classified in the Galaxy Zoo. This morphological distribution is well matched by the control sample, comprised of 35\%, 61\% and 2\% of elliptical, spiral and merging galaxies respectively \cite[see][for more details]{Riffel23_megacube}. Regarding the nuclear activity for the AGN sample, 60\% are LINER (low-ionization nuclear emission-line region) and 40\% are Seyfert, mostly type 2 (S2), with only 25 sources classified as type 1 (S1), based in BPT diagram.


\section{Measurements}
\label{sec_3}

We have used fits to the [O\,{\sc iii}]$\lambda$5007\AA\, (hereafter [O\,{\sc iii}]) emission lines to map the kinematics of the ionised gas obtained using the {\sc IFSCUBE} code \citep{daniel_IFSCUBE,Dutra_21,dutra_ifscube_22}: the continuum and underlying stellar population were obtained from fits performed in \citet{Riffel23_megacube}, available in the resulting {\sc megacubes}. The emission-line profiles were represented by one or two Gaussian curves, and the residual continuum was reproduced by a first-order polynomial function. The fits were performed only in spaxels with signal-to-noise ratio in the [O\,{\sc iii}] emission lines higher than 3, to eliminate non-significant data. The noise was calculated as the root-mean-square deviation of the continuum adjacent to the emission line.

We performed simultaneously the fit of the following emission lines for  both the AGN and control galaxies samples, using mostly a single Gaussian component for: H$\beta$, [O\,{\sc iii}]$\lambda\lambda$4959,5007, H$\alpha$, [N\,{\sc ii}]$\lambda\lambda$ 6548, 6583 and [S\,{\sc ii}]$\lambda\lambda$6716,6731. The doublet emission lines were forced to have the same kinematic properties (centroid velocities $v$ and velocity dispersion $\sigma$). We also fixed the ratios between these doublet lines according to their expected ratios: [O\,{\sc iii}]$\lambda\lambda$ 5007/4959=2.86 and [N\,{\sc ii}]$\lambda\lambda$ 6583/6548=2.94 \citep{Osterbrock_06}.

For the AGN hosts that presented a residual along the fitted [O\,{\sc iii}]$\lambda\lambda$4959,5007 profiles (mostly in the wings) higher than 2 times the noise in the neighboring continuum, we used two components in the fit: a narrow and a broader one, to fit the profile wings which were forced to have different kinematic properties, with the velocity dispersion for the broad component required to have a higher value than that of the narrow ($\sigma_b > \sigma_n$).

\begin{figure}
\begin{subfigure}{\columnwidth}
  \centering
  \includegraphics[trim = 10mm 8mm 15mm 10mm,clip,scale=0.3]{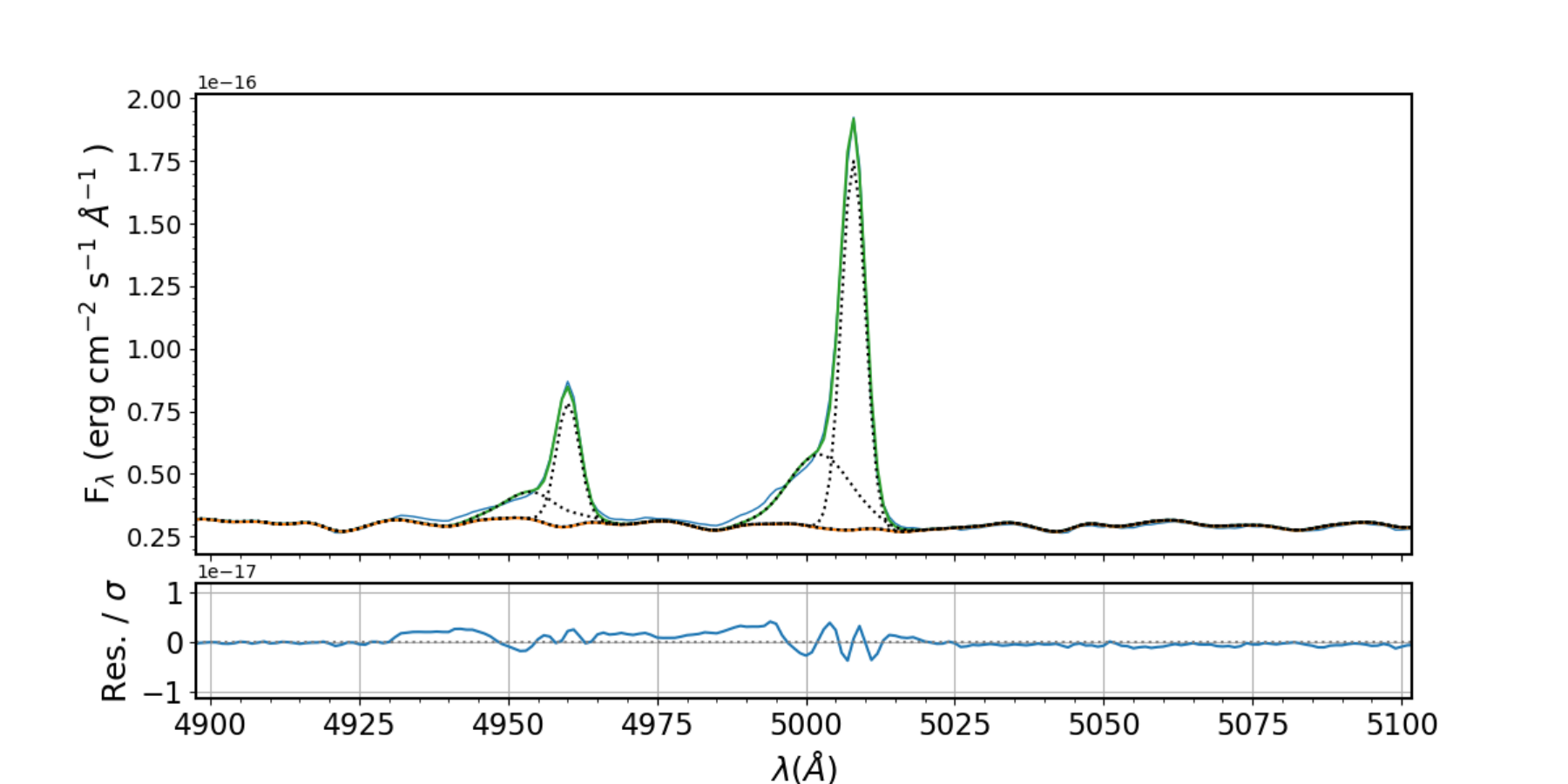}  
\end{subfigure}
\begin{subfigure}{\columnwidth}
  \centering
  \includegraphics[trim = 10mm 0mm 15mm 10mm,clip,scale=0.3]{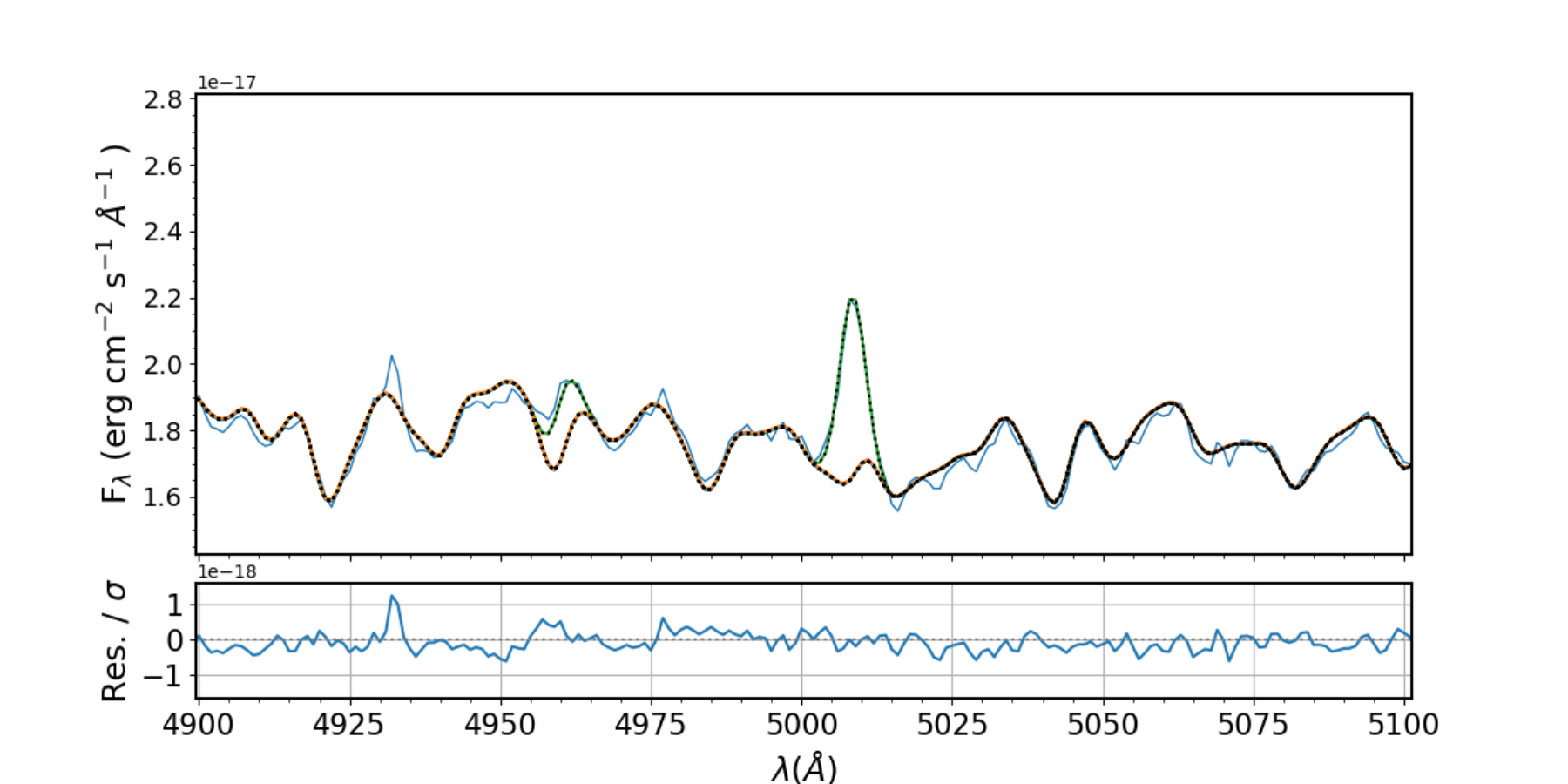}  
\end{subfigure}
\caption{Examples of the [O\,{\sc iii}] doublet emission-line fitting. \textit{Top panel}: central spaxel of the MaNGA galaxy 1-153627, using two components: narrow $+$ broad; \textit{Bottom panel:} central spaxel of the galaxy 1-25554 using one component.In this figure, the blue line represents the observed spectrum, the green line depicts the fitted spectrum, and the dashed black line represents the various components utilised in the fit.}
\label{fig:ajuste}
\end{figure}

The top panel of Figure \ref{fig:ajuste} shows an example of a fit for the central spaxel of a galaxy in which two components were required -- a narrow and a broader one, while the bottom panel shows an example of a fit in which one component was sufficient to reproduce the line profile.

In our analysis of the 293 AGN datacubes, 135 ($\approx$45\%) showed a broadening in the wings of the [O\,{\sc iii}] line profiles that required the use of two Gaussian curves to better fit the emission line, at least in the central region of the galaxy (see Fig.\,\ref{fig:ajuste}). The remaining 158 AGN ($\approx$55\%) did not present this broadening, thus we used only one Gaussian to represent the profiles in these cases.

\begin{figure*}
\includegraphics[trim = 40mm 20mm 20mm 20mm,clip,scale=0.6]{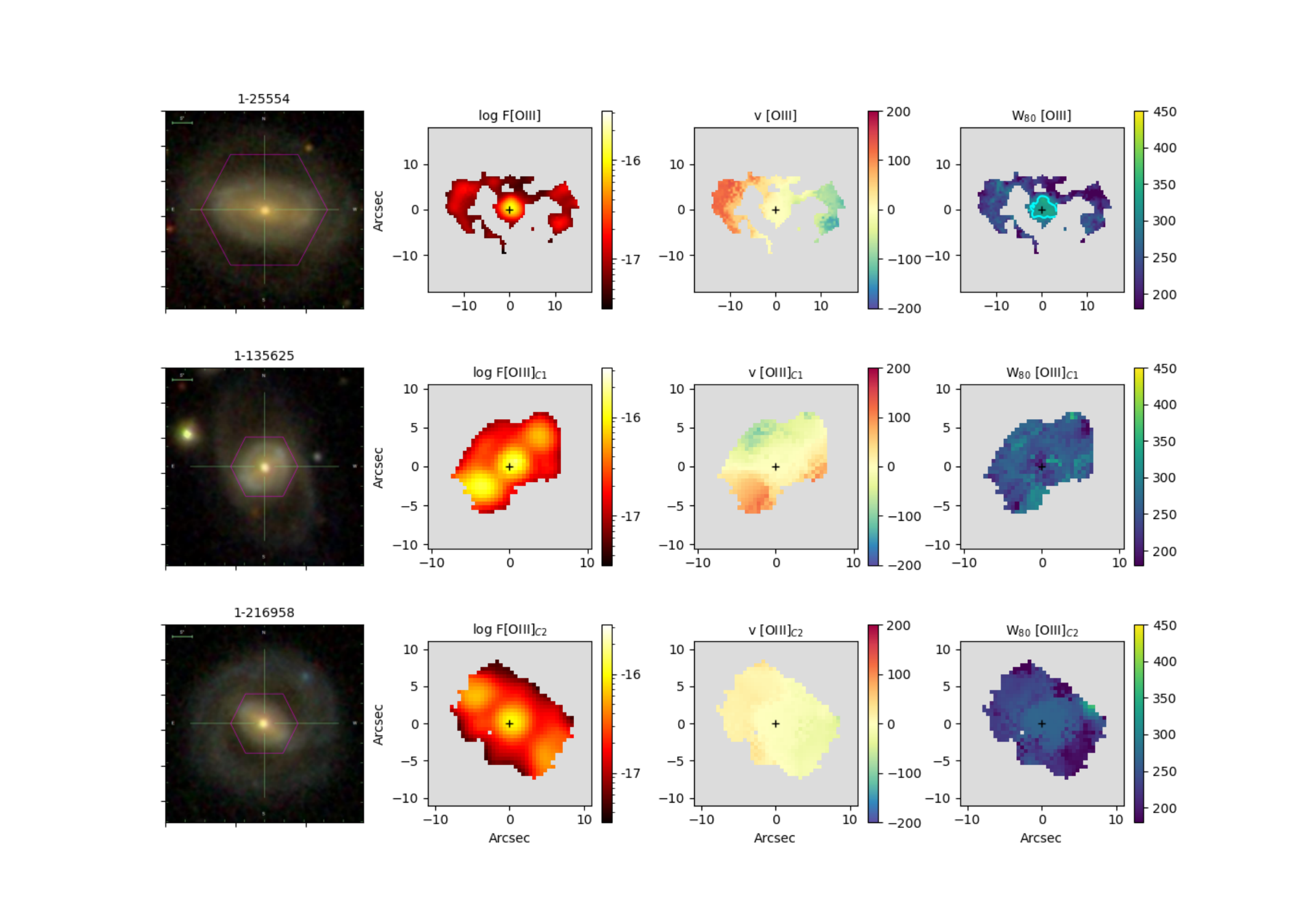}
    \caption{Maps of the AGN 1-25554 (upper row) and its control galaxies 1-135625 (middle row) and 1-216958 (bottom row). In the columns from left to right: the SDSS composite image; the [O\,{\sc iii}] flux distribution (in erg\,s$^{-1}$ cm$^{-2}$ spaxel$^{-1}$) in logarithmic scale; the [O\,{\sc iii}] centroid velocity map (in km\,s$^{-1}$); and the W$_{80}$ map (in km\,s$^{-1}$). The cyan contour in the W$_{80}$ map delimits the region with values above 315\,km\,s$^{-1}$ (see text). The grey areas identify locations where the emission-line amplitude is below 3 times the continuum noise amplitude.}
    \label{fig:mapa_1c}
\end{figure*}

\begin{figure*}
    \centering
    \includegraphics[trim = 15mm 40mm 25mm 40mm,clip,scale=0.55]{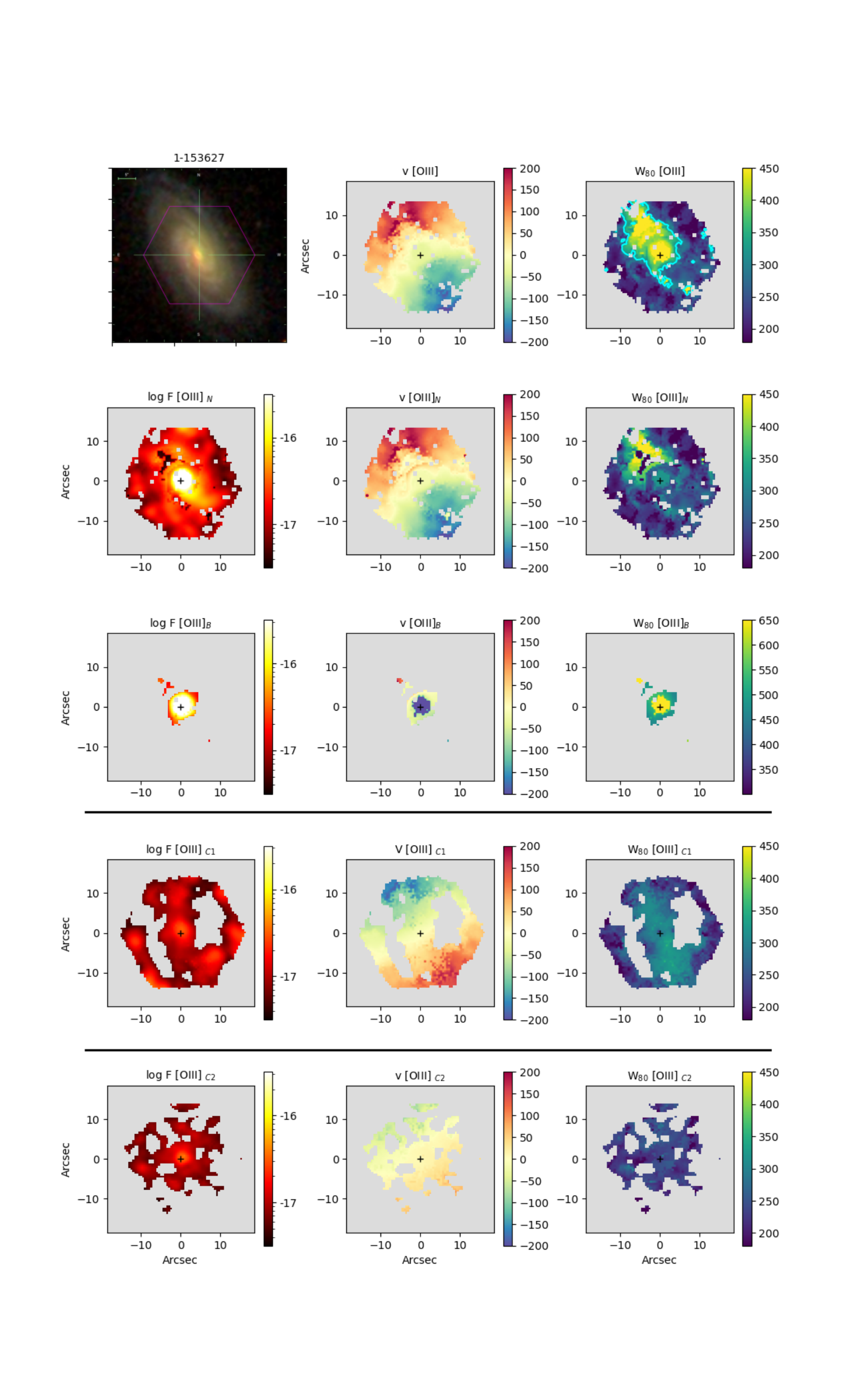}
    \caption{Flux and kinematic maps for the AGN 1-153627 (top three rows of panels) and its control galaxies 1-41828 (middle row of panels) and 1-201812  (bottom row of panels). The top row shows, from left to right, the AGN host image, the composite (broad$+$narrow) centroid velocity and W$_{80}$.
    In the second and third rows we show respectively the same properties of the narrow and broad components of the AGN, while in the bottom two rows we show them for the control galaxies. Units as in Fig. \ref{fig:mapa_1c}. The cyan contour in W$_{80}$ delimits the region with values above 315\,km\,s$^{-1}$ (see text). The grey areas identify locations where the emission-line amplitude is below 3 times the continuum noise amplitude.}
    \label{fig:mapa_2c}
\end{figure*}
 
The fits of the [O\,{\sc iii}] emission lines with Gaussian curves provided the values of the centroid velocities, corresponding to the wavelength of the Gaussian peak -- hereafter referred to as v[O\,{\sc iii}], velocity dispersion ($\sigma$) and the fluxes for both the broad and narrow components. 

\subsection{W$_{80}$ measurements}

We have used {\sc IFSCUBE} to measure the parameter W$_{80}$ which is defined as the difference between the velocities v$_{90}$ -- the velocity corresponding to the wavelength that includes 90\% of the line flux, and v$_{10}$ -- the velocity corresponding to the wavelength that includes 10\% of the line flux \citep{Liu,Harrison_2014}. The advantage of W$_{80}$ is that it is a "non-parametric" measurement, in the sense that you do not have to assume any type of profile or number of components in the fit. W$_{80}$ will measure the profile broadening closer to its wings than the velocity dispersion, thus probing the highest velocities that are usually attributed to kinetic disturbances not associated to the orbital motion in the galactic potential, such as nuclear outflows or turbulence. For a single Gaussian profile W$_{80}$ approximately corresponds to the FWHM of the emission line, or, more precisely, W$_{80}$ = 2.56$\sigma$.

Previous works have used the W$_{80}$[O\,{\sc iii}] parameter to identify regions in galaxies where a nuclear AGN significantly affects the gas kinematics as well as to trace outflows. In a previous study of MaNGA AGN, \cite{Dominika} have proposed that values of the mean W$_{80} \geq$ 500\,km\,s$^{-1}$ indicate radiatively or mechanically-driven AGN outflows. Other works argue that values of W$_{80} \geq$ 600\,km\,s$^{-1}$ are a signature of gas in unbound orbits, probably in outflow \citep[e.g.][]{Harrison_2014,Kakkad_superII,Dutra_AGNIFS}. These limits were chosen to make sure that the corresponding velocities were not due to orbital motion in the gravitational potential of the galaxy, and depend on the mass of the galaxies. Even for the most massive galaxies, velocities due to orbital motion cannot reach above 500\,--\,600\,km\,s$^{-1}$ and thus can only be due to other "non-gravitational" kinematics.

For lower mass galaxies, such values should be lower, and, in our case, we can use as reference the control galaxies that have been matched to the AGN hosts in galaxy mass, distance, morphology and stellar kinematics, as previously discussed. We thus can use the control sample to establish the threshold W$_{80}$ value above which the gas kinematics is affected by disturbances due to the AGN.

In the comparison between AGN and controls, we use a ``weighted mean"  $\langle$W$_{80}\rangle$, in which the weight is the corresponding line flux at each spaxel:
\begin{equation}
    \langle W_{80} \rangle = \frac{\sum (W_{80} \,F[O\,{\sc III}])}{\sum F[O\,{\sc III}] }
    \label{eq:w80_medi}
\end{equation}
This weighting attributes a greater contribution to the mean from the spaxels that have higher fluxes and a lower contribution from spaxels with lower fluxes, that are usually more distant from the nuclear region, therefore are less affected by the AGN and are also noisier.

\subsection{Luminosity measurements}
\label{sec:L_bol}

We have characterised our sample [O\,{\sc iii}] luminosity $L_{\rm [O\:III]}$, by that measured within the central 3\arcsec x 3\arcsec\, square (just above the typical MaNGA PSF\,--\,Point Spread Function -- diameter). This is also the diameter of the central fiber of the MaNGA IFU, with 3\arcsec corresponding on average to 2.75\,kpc at the galaxies of our sample. We decided to adopt this value to characterise our sample to keep consistency with our previous studies, in which we used the same procedure. We have then corrected $L_{\rm [O\:III]}$ by reddening using the relation of \cite{Lamastra}. 

The bolometric luminosity L$_{bol}$ was obtained through the relation between L$_{bol}$ and the reddening-corrected $L_{\rm [O\:III]}$ using the equation \citet{Trump_2015}:
\begin{equation}
    \frac{L_{bol}}{10^{40}\,erg\,s^{-1}} = 112 \left( \frac{L_{[O{\sc III}]}}{10^{40}\,erg\,s^{-1}}\right)^{1.2}
\end{equation}


\section{Results}
\label{sec_4}
\subsection{Flux and kinematic maps}

In Figure \ref{fig:mapa_1c} we show the flux and kinematic maps for the AGN host galaxy 1-25554 (top panel) that required only a single Gaussian to fit the [O\,{\sc iii}] profile, compared to those of its two control galaxies (middle and bottom rows of panels). For each source we show the SDSS composite image in the first column. Figure \ref{fig:mapa_1c} shows that the velocity fields $v[O\,{\sc III}]$ of both the AGN and the first control C1 present a rotation pattern, while for the control C2 the velocity is close to zero everywhere, probably due to a somewhat more face on orientation of the galaxy. Regarding the W$_{80}$, the AGN values are higher than those of its controls within $\approx 3-4^{\prime\prime}$ from the nucleus. We have concluded that the W$_{80}$ value that separates ``pure" rotation (orbital motion in the galaxy potential) from the possible contribution from outflows is 315\,km\,s$^{-1}$, as discussed below, which is delimited by a cyan contour in Fig. \ref{fig:mapa_1c}.

In Figure \ref{fig:mapa_2c} we show the maps for the AGN host galaxy 1-153627 (top three rows of panels) that required two Gaussian curves to fit [O\,{\sc iii}], compared to those of its control galaxies (middle and bottom rows of panels). In the upper row we show the maps for the two components -- narrow and broad, combined, which we call composite maps (measurements performed using the narrow and broad components added together). In the second row we show the maps for the narrow component only and in the third row the maps for the broad component only. 
The AGN composite velocity field is dominated by that of the narrow component in the outer regions of the galaxy, showing redshifts to the northeast and blueshifts to the southwest, while the broad component is confined to the inner $\approx$ 3$^{\prime\prime}$ radius and shows only blueshifts of up to $\approx -200$\,km\,s$^{-1}$.  Comparing the values of W$_{80}$ from the AGN to those of its control galaxies we see higher values for the AGN in the composite maps in a patch covering from the nucleus to the border of the galaxy to the northeast. Outside the nucleus, the W$_{80}$ values of the narrow component are also higher than those of the controls, suggesting that some kinematic disturbance from the AGN is also present in the narrow component. For the broad component, the highest values of W$_{80}$ reach up to 650\,km\,s$^{-1}$.

\begin{figure*}
    \centering
    \includegraphics[trim = 45mm 0mm 55mm 0mm,clip,width=1\textwidth]{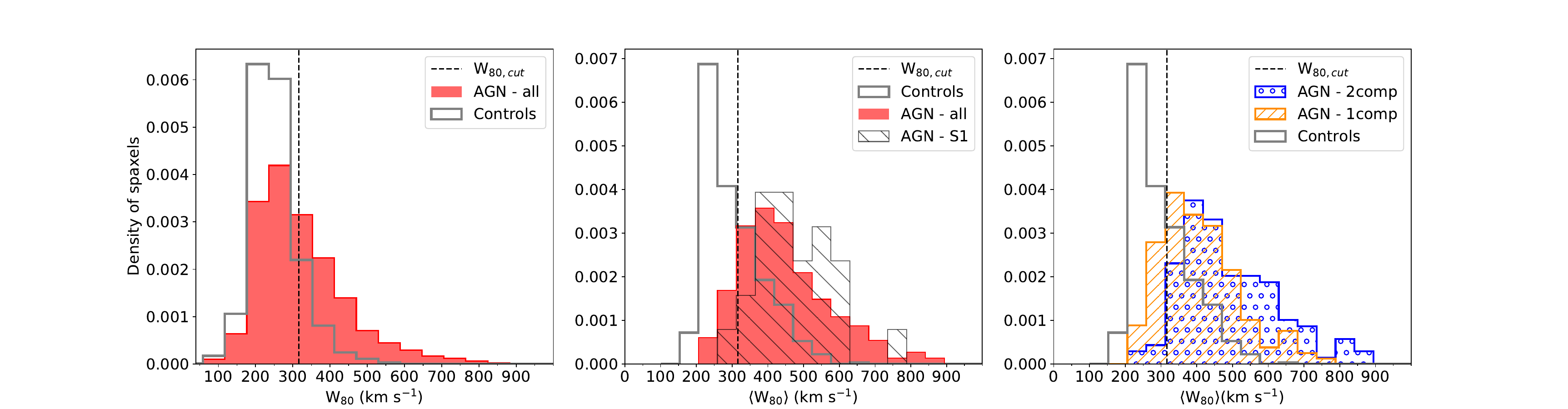}
    \caption{\textit{Left panel:} Distribution of W$_{80}$ measurements for all spaxels of the AGN and control samples. AGN are shown in red and controls in grey. The dashed line shows the adopted cut value W$_{80,cut}\approx 315$\,km\,s$^{-1}$ (see sec.\ref{kdr}). \textit{Central panel:} Weighted mean $\langle$W$_{80}\rangle$ for each galaxy, where again the controls and AGN are shown in grey and red, respectively and the hatched histogram shows the 25 Sy 1 galaxies of the sample. \textit{Right panel:} $\langle$W$_{80}\rangle$ values for the AGN with only a narrow component in [O\,{\sc III}] (hatch orange histogram), as compared with those with two components (dotted blue histogram), and control galaxies (empty grey histogram).}
    \label{fig:all_hist}
\end{figure*}

The maps for the remaining objects are included as supplementary material. The behaviour is similar to that seen in Figs. \ref{fig:mapa_1c} and \ref{fig:mapa_2c}: the velocity field v[O\,{\sc iii}] frequently shows a rotation pattern that is similar for the AGN and controls but the W$_{80}$ map almost reach higher values for the AGN than for the controls within the inner few arcseconds, corresponding to typical distances at the galaxies of $\approx$ 3\,kpc. But there are cases, as in Fig.\,\ref{fig:mapa_2c}, where higher W$_{80}$ values are seen for the AGN to larger distances, and up to the border of the [O\,{\sc iii}] emission-line map. This is further discussed in the next sections via the measurement of the associated Kinematically Disturbed Region (KDR). 

\subsection{W$_{80}$ values}
In order to compare the values of W$_{80}$ for the AGN and controls, we have built an histogram of the values from all spaxels of all galaxies, shown in the left panel of Figure~\ref{fig:all_hist}: the distributions median values are 323\,km\,s$^{-1}$ and 252\,km\,s$^{-1}$ for the AGN and control galaxies, respectively. The KS test gives a p-value $< 10^{-116}$ when comparing the two distributions, revealing that the kinematic properties of the two samples are distinct. The black dashed line in the histogram marks the value of W$_{80,cut}=315$\,km\,s$^{-1}$ described in Section\,\ref{kdr}.

The central panel of Figure~\ref{fig:all_hist} shows the distribution of the weighted mean $\langle$W$_{80}\rangle$ values (equation\,\ref{eq:w80_medi}) obtained for each galaxy of the two samples. For the AGN hosts that have both narrow and broad components the $\langle$W$_{80}\rangle$ values are those from the composite profiles.
This distribution has a median value of 453\,km\,s$^{-1}$ for the AGN (in red) and 306\,km\,s$^{-1}$ for the controls (in gray). The AGN have higher values on average, and a broader distribution of $\langle W_{80}\rangle$, reaching up to $\approx$900\,km\,s$^{-1}$, while the the values for the controls reach $\approx$ 500\,km\,s$^{-1}$ for very few sources. 
We also include in the central panel of Fig.\,\ref{fig:all_hist} a hatched histogram to show the distribution of $\langle$W$_{80}\rangle$ values for the 25 type 1 AGN of our sample, showing that their values are similar to those of the dominant type 2 sources.

The right panel of Figure~\ref{fig:all_hist} shows a comparison of the $\langle$W$_{80}\rangle$ values separately for the narrow (in hatched orange histogram) and broad components (in dotted blue histogram) for the AGN that have these two components (45\% of the sample) with those of the control galaxies. Although the narrow component values overlap with those for the controls, the K-S test show low p-values of about 10$^{-4}$ when comparing the two distributions, indicating that they are somewhat different, with the AGN showing slightly higher values. The comparison of the  $\langle$W$_{80}\rangle$ distribution of the broad component with those of the control galaxies results in a much lower p-value, of 10$^{-39}$, as expected and clearly shown in the right panel of Figure\,\ref{fig:all_hist}. The gas kinematics of the narrow component of the AGN -- although not completely equal to that of the control galaxies -- is much more similar to it than that of the broad component. This narrow component can thus be mostly attributed to orbital motion of the gas in the disks of the galaxies. On the other hand, the kinematics of the broad component is characterised by higher velocities than the controls, indicating that it is not due to gravitational motions, being usually attributed to outflows \citep[e.g.][]{Bruno_21}.


\section{Discussion}\label{sec:5}

\subsection{Larger W$_{80}$ values for AGN than controls}
As shown in Figure~\ref{fig:all_hist} and discussed above, the W$_{80}$ and $\langle$W$_{80}\rangle$ values are usually larger for the AGN than for the controls within the inner $3-4$\arcsec, and in many cases extending out into the galaxy, up to the border of the [O\,{\sc iii}] emission distribution. 

We attribute these larger W$_{80}$ values in the AGN to kinematic disturbances in the gas produced by the AGN via radiation pressure, heating and/or nuclear outflows \citep{rogemar_22,Dominika,Dutra_AGNIFS,Venturi_2021,rogerio_riffel_2023}. 
This result is similar to that found by \citet{Alice_paperV} using a sub-sample of the AGN and controls of our study, and a somewhat different method. They found that the AGN hosts had higher values of residual ionised gas velocities (centroid velocities relative to systemic) and velocity dispersion than the controls, especially for AGN with luminosities greater than $L_{\rm [O\:III]}$ $\geq 3.8 \times 10^{40}$ erg s$^{-1}$, in agreement with the  results of our study.

Larger values of W$_{70}$ -- difference between the 85th percentile and 15th percentile velocities of the fitted line profile -- have similarly been observed by \cite{Venturi_2021} associated to outflows and ionisation cones, but, instead of the enhanced values being observed along the outflows, they are observed perpendicularly to them. We thus consider that W$_{80}$ can also be probing this kinematic effect.

We note that there is no difference in the distribution of values for W$_{80}$ between S1 and S2 galaxies. Within the framework of the Unified Model of AGN \citep{Antonucci85b}, in which the AGN in S1 galaxies would be observed more ``pole-on" and in S2 more ``edge-on", this similarity suggests that the kinematic disturbances we are probing do not show a preferred orientation, and are dominated by an approximately ``isotropic" disturbance, e.g., an ``spherical outflow", or enhanced velocity dispersion due to turbulence in the gas, such as those found perpendicular to the ionisation axis or radio jet in recent studies \citep{Riffel_2015,Dutra_21,Bruno_21,2021_Rifel_chemicalAbundance}.  On the other hand, we do find an excess in the percentage of type 1 showing a broad component relative to that in type 2: -- 60\% for type 1  vs. 44\% for type 2 sources, that supports that the broad component can be due to an outflow seen more ``pole on" in the type 1 sources.

\begin{figure*}
    \centering
    \includegraphics[trim = 10mm 10mm 20mm 10mm,clip,width=0.7\textwidth]{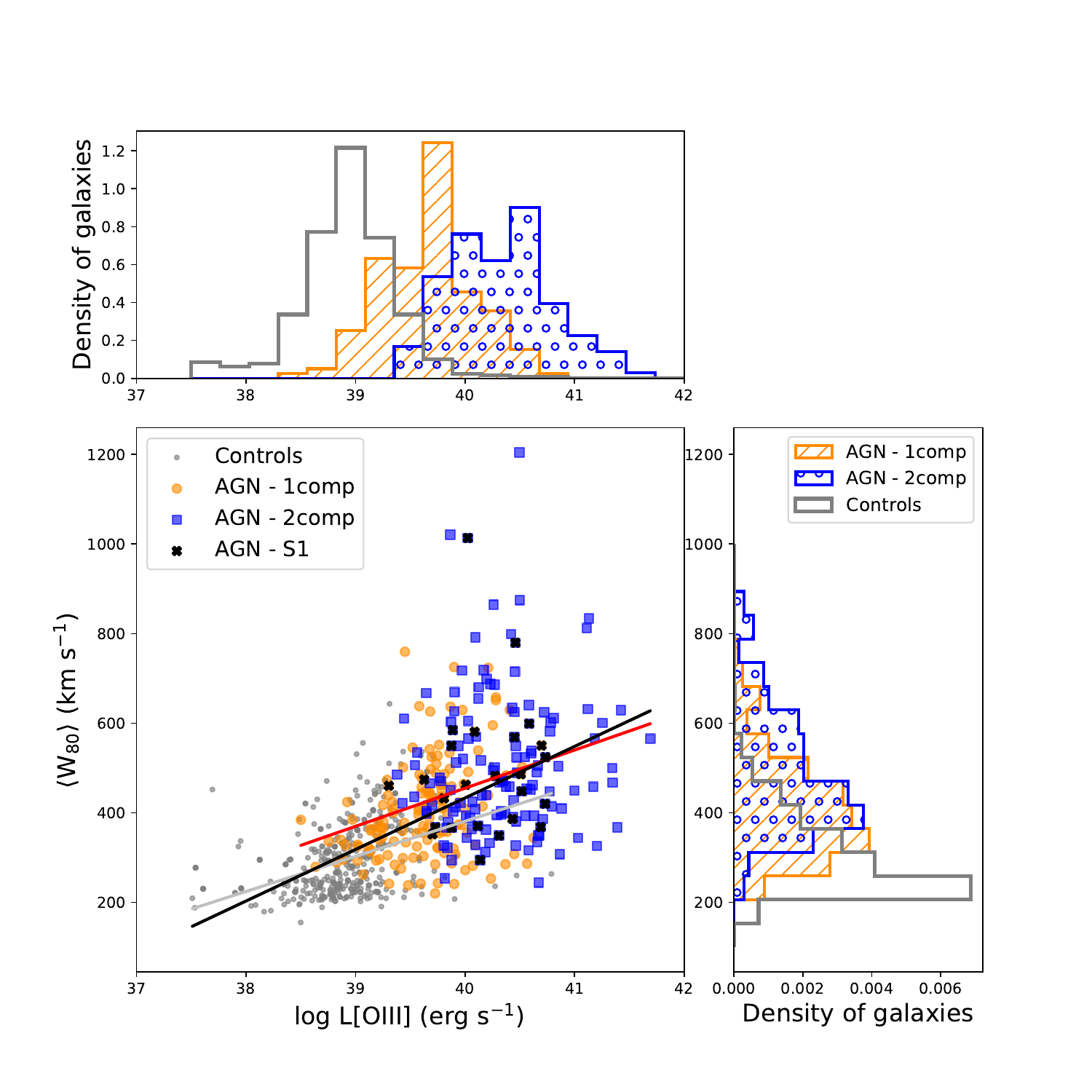}
    \caption{Relation between $\langle$W$_{80}\rangle$ and $L_{\rm [O\:III]}$. The AGN sample is shown in orange (profile showing only one component) and in blue (two components), the controls in gray.  The S1 sources are identified as black crosses -- most of them superimposed on the blue circles (two components). The lines are linear regressions for the AGN sample (red), controls (gray), and for the AGN and controls combined (black).}
    \label{fig:lX80}
\end{figure*}

In Figure~\ref{fig:lX80} we show the relation between $\langle$W$_{80}\rangle$ and $L_{\rm [O\:III]}$ for the AGN and control samples. We find a positive correlation between these two properties, not only for the AGN (orange circles and blue squares) but also for the the control galaxies (gray dots). An even stronger correlation is found for the combined control and AGN samples. Furthermore, we plot separately in Figure\,\ref{fig:lX80} the two AGN data samples, those that have only one Gaussian component in [O\,{\sc iii}], in orange circles, and those that have a narrow and a broad component in blue squares. As expected the AGNs that have a broad component tend to present the highest values of $\langle$W$_{80}\rangle$ and $L_{\rm [O\:III]}$.

This correlation confirms previous results that the AGN luminosity plays a dominant role in introducing kinematic disturbances and outflows in the host galaxies \citep{Dominika,Kakkad_superII}.

We fit the AGN data in Fig.\ref{fig:lX80} by a linear equation and find:
\begin{equation}
     \langle W_{80,AGN}\rangle \,=  85.09 \,log(L_{[O{\sc III}]}) - 2948.92
\end{equation}
This linear regression is shown as a red line in Figure~\ref{fig:lX80}. The corresponding Spearman correlation coefficient is $\rho$= 0.36 with associated p-value = $10^{-10}$, supporting a weak correlation between the gas kinematic disturbance traced by W80 and the AGN luminosity. Unexpectedly, this effect seems also to exist for the control sample, for which we find a similar relation:
\begin{equation}
     \langle W_{80,C} \rangle \,=  78.83 \,log(L_{[O{\sc III}]}) - 2766.76,
\end{equation}
with $\rho$= 0.39 and associated p-value = $10^{-26}$. This linear relation is shown in grey in Figure~\ref{fig:lX80}.

Considering the AGN and control samples together, we obtain a similar relation but with a better correlation:
\begin{equation}
    \langle W_{80,AGN+C} \rangle \,= 114.11 \,log(L_{[O{\sc III}]}) - 4129.36
\end{equation}
with $\rho$=0.66, with a p-value = 10$^{-99}$. This regression is shown as a black line in Figure~\ref{fig:lX80}.

This result is similar to that found by \citet{ilha_paperIII}, in which the velocity dispersions of the controls combined with those of the AGN are correlated with $L_{\rm [O\:III]}$. This is probably due to the fact that the controls have lower $L_{\rm [O\:III]}$ than the AGN, as well as lower velocity dispersions and consequently lower W$_{80}$, introducing a better correlation with $L_{\rm [O\:III]}$ when combined with the higher AGN values.
Alternatively, there may be some weak AGN contribution in at least part of the control galaxies. This can be understood as due to the fact that the threshold between LINERs and LIERs (that can be excited by hot evolved stars) only indicates that the LIERs {\it may not} be actual AGN, but could be, as there should be no lower limit for an AGN luminosity. Alternatively, irrespective of the ionization source -- AGN or hot evolved stars -- the luminosity of the central source has a kinematic effect on the gas.

\subsection{The difference in $\langle W_{80}\rangle$ between AGN and controls}

As we have matched each AGN host to two control galaxies, we can also compare the $\langle W_{80}\rangle$ values of each AGN to those of its control galaxies by calculating the difference $\Delta_{W_{80}}$ between the AGN value and the average of the values of its two controls:
\begin{equation}
    \Delta_{W_{80}} = \langle W_{80}\rangle_{AGN} - \frac{\langle W_{80}\rangle_{C1}+\langle W_{80}\rangle_{C2}}{2},
\end{equation}

\begin{figure}
    \centering
    \includegraphics[trim = 27mm 0mm 35mm 15mm,clip,width=\columnwidth]{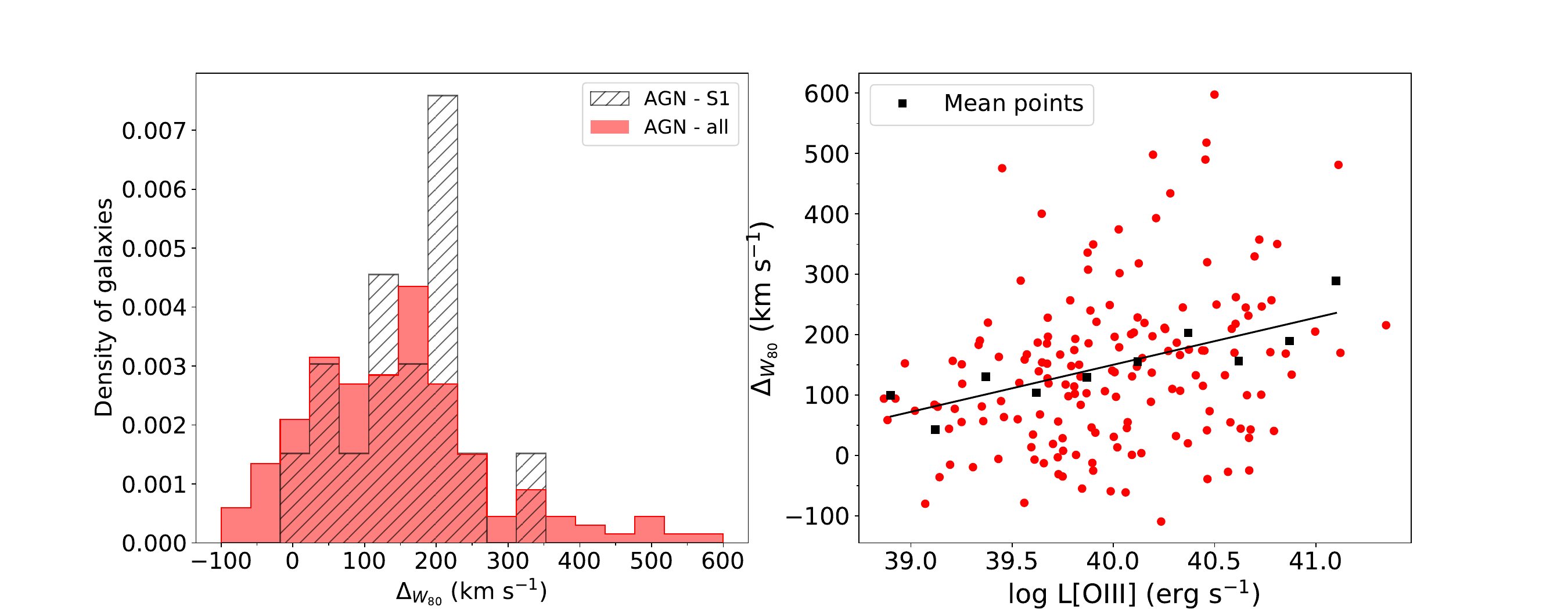}
    \caption{\textit{Left panel:} Distribution of values of $\Delta_{W_{80}}$ -- the difference between the $\langle$W$_{80}\rangle$ values of the AGN and their two controls. The black hatched histogram shows the values for the 25 S1 AGN. \textit{Right panel} Relation between $\Delta_{W_{80}}$ and $L_{\rm [O\:III]}$; the black square show mean values within bins of luminosity, while the solid black line shows a linear regression to these points.}
    \label{fig:delta}
\end{figure}

In the left panel of Figure~\ref{fig:delta} we show an histogram of the $\Delta_{W_{80}}$ values, that are in the range $-$109--598\,km\,s$^{-1}$, showing a mean value of 143\,km\,s$^{-1}$. Only 20 AGN (7\%) of our sample do not have $\langle W_{80}\rangle$ values higher than those of their controls. The black hatched histogram in Figure~\ref{fig:delta} refers to the 25 S1 AGN in our sample, all showing $\langle W_{80}\rangle$ values higher than those of their controls, other than that, they follow the same distribution of the type 2 AGN.

The right panel of Figure~\ref{fig:delta} shows the relation between $\Delta_{W_{80}}$ and $L_{\rm [O\:III]}$. We calculated the mean $\Delta_{W_{80}}$ values within logarithmic luminosity bins of 0.25 (in units erg\,s$^{-1}$), showing these mean values as black square, then performed a linear regression to these mean values as a function of $L_{\rm [O\:III]}$ with the following equation
\begin{equation}
    \Delta_{W_{80}} = 78 \,log(L_{[O{\sc III}]}) - 2969.99,
\end{equation}
which is shown by the black line in Fig.\,\ref{fig:delta}. These mean values show a strong correlation with $L_{\rm [O\:III]}$ with a Spearman's rank correlation coefficient of 0.91 and a p-value$=10^{-4}$.

\subsection{The role of the bulge}
\label{bojo}

\begin{figure*}
    \centering
    \includegraphics[trim = 45mm 0mm 50mm 5mm,clip,width=1\textwidth]{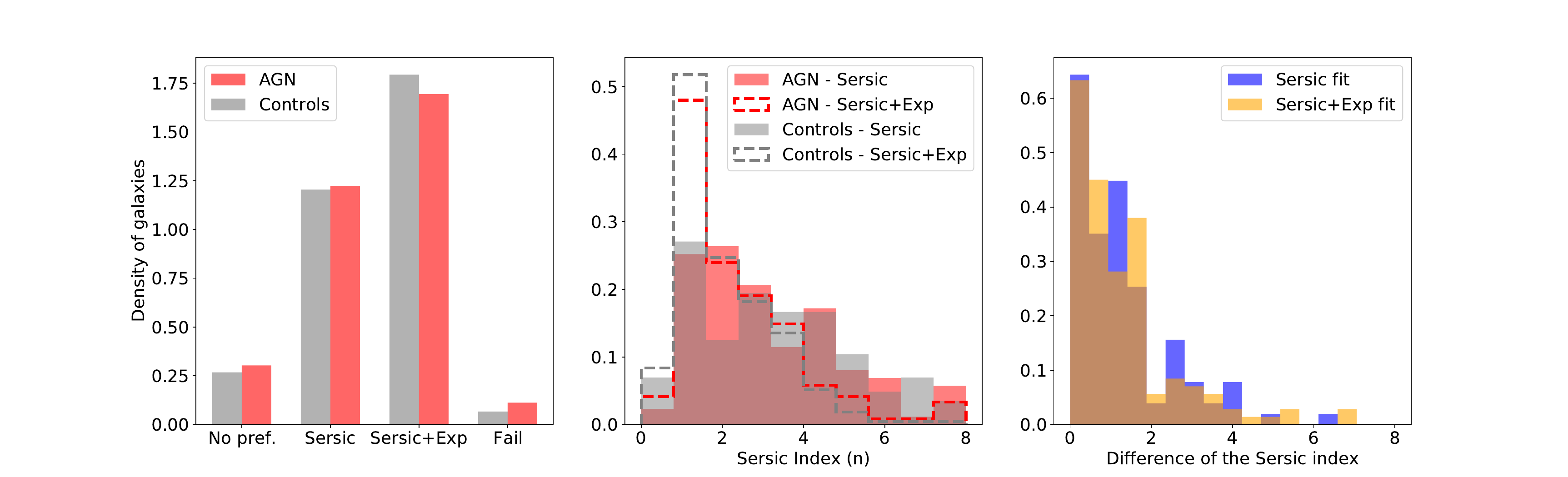}
    \caption{\textit{Left panel:} Distribution of AGN and controls according to the types of their photometric profiles: Sersic only, Sersic$+$Exponential, no preference (both types provide similar fits -- No pref.) or ``Fail" (the fit fails).  \textit{Central panel:} Distribution of the Sersic indices for the bulge for both profile types. \textit{Right panel}: Distribution of the difference in the Sersic index values between the AGN and the average value of their controls. }
    \label{fig:hist_sersic}
\end{figure*}

Early studies of the NLR gas kinematics have pointed out the influence of the gravitational potential of the galaxy, argued to drive the correlation between the [O{\sc iii}] line profile FWHM and virial parameters \citep{Whittle_1992a,Whittle_1992b,Whittle_1992c}. In particular, \citet{Veilleux_1991b,Veilleux_1991c} found a positive correlation between the [O{\sc iii}] line width and the luminosity of the galactic bulge, with early-type galaxies presenting a tendency to show broader line profiles than late-type ones, indicating the importance of the host galaxy bulge in the low-velocity gas dynamics in the NLR.

We have used a control sample matched via the host galaxy properties in order to avoid differences in the gas kinematics that could be due to different properties of the bulge. This should assure that the difference between the AGN and their controls is due to the AGN activity and not to the properties of the bulge. As already discussed in Sec.\,\ref{sec_2}, when presenting the sample, Figure \ref{fig:hist_sample}-- third panel, shows that the distribution of the stellar velocity dispersion for the two samples is very similar (p-value 0.59), indicating that the galaxy gravitational potential is similar for AGN and controls, and that the difference of properties between AGN and controls should not be driven by the bulge.

In any case, we investigate also the photometric properties of the bulges of the galaxies of our sample in order to further check if the AGN and controls are matched also in terms of these properties. We use the MaNGA PyMorph DR17 photometric catalog\footnote{\url{https://data.sdss.org/datamodel/files/MANGA_PHOTO/pymorph/PYMORPH_VER}} to obtain the bulge photometric parameters for both the AGN and control galaxies. In particular, we use the results of the fits of Sersic and Sersic$+$Exponential laws to their 2D surface brightness distributions of the final MaNGA DR17 galaxy sample.

In the left panel of Figure \ref{fig:hist_sersic} we show the distribution of the AGN as compared to that of the controls in terms of their photometric profile types: (1) profiles well reproduced by a Sersic law only; (2) profiles that are composite, comprising a Sersic plus an exponential profile; (3) profiles that are well fit by both types of profiles (no preference), and (4) profiles for which the fit has failed. This figure shows that the distributions of profile types are similar for the AGN and controls.

In the central panel of Figure \ref{fig:hist_sersic} we show the distributions of the Sersic index values (\textit{n}), in the range $0-8$ for the AGN and controls, considering the two different photometric profile types: Sersic and Sersic$+$Exponential. For the galaxies well fitted with a Sersic profile (filled histograms), the distributions for the AGN and controls are similar, with a mean value of $n=3.1$ for the AGN and $n=3.2$ for the controls. The same applies for galaxies that have a composite Sersic$+$Exponential profile: the mean value for the AGN is $n=2.3$, and for the controls is $n=2$. Again, this figure shows that the distributions are similar for the AGN and controls.

In the right panel of Figure \ref{fig:hist_sersic} we now check if the AGN and controls are well paired in terms of their Sersic index.
In this figure, we show the distribution of the differences in the Sersic index between each AGN and their respective controls (value for the AGN minus the average of the values for the controls). We find that for approximately 80\% of the AGN the difference is smaller than 2, showing that, in most cases. the AGN and controls are indeed well paired according to their bulge photometric properties.

\begin{figure*}
    \centering
    \includegraphics[trim = 25mm 0mm 35mm 15mm,clip,width=1\textwidth]{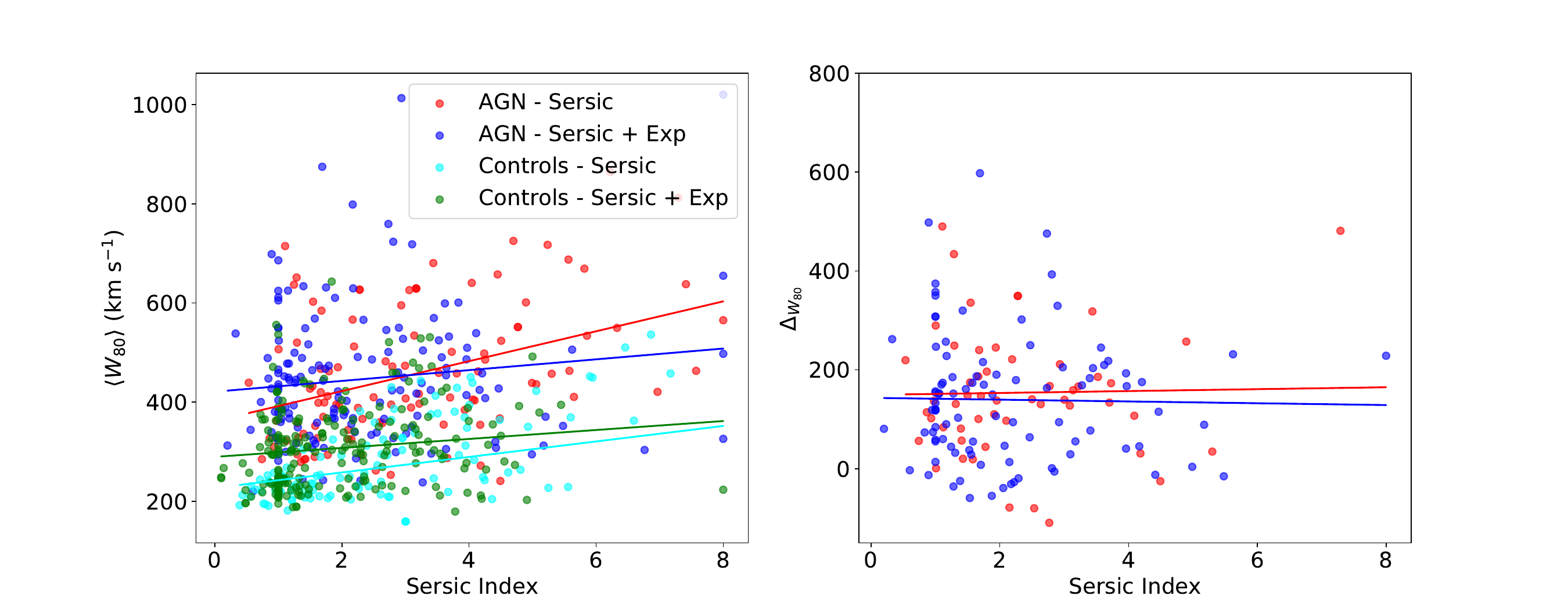}
    \caption{\textit{Left panel}: Relation between $\langle$W$_{80} \rangle$ and the Sersic index; \textit{right panel}: relation between $\Delta_{W_{80}}$ and Sersic index. The colours points are described in the legend of the figure. The solid lines represent the linear regressions following the colours of the corresponding points, the fit parameters are described in Tab. \ref{tab:parametros}.}
    \label{fig:n_sersicVSvelocidades}
\end{figure*}

We now investigate the relation between the $\langle W_{80} \rangle$ values and the Sersic index \textit{n} of the galaxy bulges. In the left panel of Figure \ref{fig:n_sersicVSvelocidades}, we present this relation between for AGN and controls for both types of photometric profiles, Sersic only and Sersic$+$Exponential. We find a weak correlation in the case of the Sersic profile for the AGN (Spearman rank correlation coefficient $\rho=0.40$) and stronger for the controls, with $\rho=0.58$. The increase in the $\langle W_{80} \rangle$ is nevertheless small, suggesting just a small role of the bulge in the gas kinematics. In the case of the composite photometric profiles, Sersic$+$Exponential, there is no correlation for the AGN and only a marginal one for the controls, as listed in Table \ref{tab:parametros}. This table also lists the Spearman rank correlation coefficients $\rho$ and p-values for the relation between the velocity dispersion $\sigma$ vs. the Sersic indexes (figure not shown), for which we find some correlation only for the Control sample with the Sersic only profile.

Finally, we test if the higher $\langle W_{80} \rangle$ values that we have found in AGN relative to controls are related to the galaxy gravitational potential, by plotting in the right panel of Figure \ref{fig:n_sersicVSvelocidades}, $\Delta_{W80}$ vs. the Sersic \textit{n} indexes for both the Sersic only and Sersic$+$Exponential profiles. The correlation coefficients and p-values listed in Table \ref{tab:parametros} show that there is no correlation at all, indicating that the AGN excess in $\langle W_{80} \rangle$ relative to the control galaxies is not related to theses indexes, further supporting that this excess is not related to the bulge. This result supports that the excess is due to kinematic disturbances produced by the AGN.

\begin{table}
\caption{Values of: the Spearman’s rank correlation coefficient $\rho$ and p-value together with slope and intersection for the linear fits shown in Figure \ref{fig:n_sersicVSvelocidades} for $\langle W_{80} \rangle$ and $\Delta_{W80}$ as a function of the Sersic index. Also shown are parameters for similar relations obtained for the velocity dispersions $\sigma$[O{\sc iii}] instead of $\langle W_{80} \rangle$ as a function of the Sersic index.}
\label{tab:parametros}
\begin{tabular}{lcccc}
\hline \hline
          \textbf{}          & \textbf{$\rho$} & \textbf{p-value} & \textbf{Slope} & \textbf{Intersection} \\ 
          & & & &  (km\, s$^{-1}$) \\
          \hline \hline
          \textbf{AGN} \\ \hline
          
          $\langle W_{80} \rangle$ vs. Sersic       & 0.40                & 1.44 e-05       & 30.2          & 361.8   \\
          $\langle W_{80} \rangle$ vs. Sersic + Exp  & 0.05                & 0.52            & 10.8          & 421.1   \\
          $\sigma$ vs. Sersic        & 0.15                & 0.06            & 8.9           & 173.8   \\
          $\sigma$ vs. Sersic + Exp  & 0.09                & 0.23            & 6.3           & 176.0   \\
          $\Delta_{W_{80}}$  vs. Sersic         & 0.01                & 0.92            & 1.92          & 149.4   \\
          $\Delta_{W_{80}}$ vs. Sersic + Exp & 0.04                & 0.70            & -1.8          & 143.3   \\ \hline
         \textbf{Controls}  \\ \hline
         $\langle W_{80} \rangle$ vs. Sersic       & 0.58                & 1.13e-11        & 15.6          & 226.8   \\
          $\langle W_{80} \rangle$ vs. Sersic + Exp & 0.20                & 0.03            & 9.0           & 289.5  \\
          $\sigma$ vs. Sersic      & 0.46                & 1.61e-08        & 5.0           & 91.1    \\
          $\sigma$ vs. Sersic + Exp  & 0.17                & 0.06            & 2.4           & 107.3   \\ \hline \hline
          
\end{tabular}
\end{table}


\subsection{The Kinematically Disturbed Region – KDR}
\label{kdr}

Previous studies in the literature have found that the extent of the (extended) narrow line region (ENLR) correlates with the AGN luminosity and ranges from a few hundred parsecs in low-luminosity AGN \citep{Jana_paperIV,Alice_paperV,Meena_2023} to several kpc in the most luminous sources \citep{Bruno_21}. \citet{fischer_18}, using HST images and long-slit spectroscopy have found that the extent of the outflows are, on average, 22\% of the extent of the NLR in nearby type 2 quasars. We now investigate in our sample the relation between the extent of the region where the gas is disturbed by the AGN and that of the ENLR -- the region where the gas is ionised by the AGN, as follows.

We define the kinematically disturbed region (hereafter KDR) as the region where the AGN significantly affects the gas kinematics (e.g. through AGN radiation and winds). As discussed above, we have searched for disturbances in the ionised gas kinematics in the [O\,{\sc iii}] emission-line profile wings, that will be more extended if there is an outflow component besides the orbital motion in the galaxy, using for this the W$_{80}$ parameter. Recent works using the [O\,{\sc iii}]5007 line to map the gas kinematics have argued that values of W$_{80}>$\,600\,km\,s$^{-1}$ are associated with ionised gas outflows in quasars and in high luminosity AGN \citep{Kakkad_superII,Dutra_AGNIFS}. Using a low-luminosity AGN sample from MaNGA, \cite{Dominika} proposes a lower value of W$_{80}>$500\,km\,s$^{-1}$, as indicating the presence of outflows in the [O\,{\sc iii}] kinematics.  

In the present study, we can use our AGN and control samples W$_{80}$ distributions to look for this W$_{80}$ ``threshold" value that characterises the KDR in our low-luminosity sample. We have thus defined W$_{80,cut}$ as the minimum value that can denote the presence of a disturbance or non-orbital motion in the ionised gas kinematics. Figure~\ref{fig:all_hist} shows the distribution of W$_{80}$ for the control galaxies, that we assume is due to orbital motion in the galaxy. We adopt its mean value plus its standard deviation as the minimum value above which there is impact in the gas kinematics from the AGN: W$_{80,cut}$ = 252 + 63 $\cong$ 315 km\,s$^{-1}$, where  252\,km\,s$^{-1}$ is the mean and the 63\,km\,s$^{-1}$ is the standard deviation of the W$_{80}$ distribution for the controls. In the histograms of Fig.\ref{fig:all_hist} the dashed vertical line marks this value of W$_{80,cut}$.

We thus adopt the KDR as the region delimiting spaxels with W$_{80}>$W$_{80,cut}$ and are classified as AGN or composite in the BPT diagram. We also define the region ionised by the AGN, the ENLR, as that corresponding to spaxels classified as AGN or composite in the BPT diagram. We then performed the fit of an ellipse to the borders of the KDR and ENLR regions to obtain their extent, R$_{ENLR}$ and R$_{KDR}$, defined as the largest distance between the center of the galaxy and the fitted ellipse. About 6.5\%  of the AGN sample does not have W$_{80}$ values higher than W$_{80,cut}$, as shown in Figure~\ref{fig:delta}, thus R$_{KDR}$ could not be determined for these.

We note that in the right histogram of Figure\,\ref{fig:lX80} the AGN that have two components (45\% of them) in the [O{\sc iii}] emission line, show $\langle W_{80} \rangle$ values higher than W$_{80,cut}$ in about 95\% of the objects. On the other hand, this percentage is 76\% for the AGN that have one component only. As the broad component is considered a strong indicator of AGN outflows,  even though being restricted to circunmnuclear regions less extended than the KDR, this high percentage supports the identification of $\langle W_{80} \rangle > W_{80,cut}$ as a tracer of AGN kinetic feedback.

\begin{figure*}
    \centering
    \includegraphics[trim = 45mm 0mm 50mm 0mm,clip,width=1\textwidth]{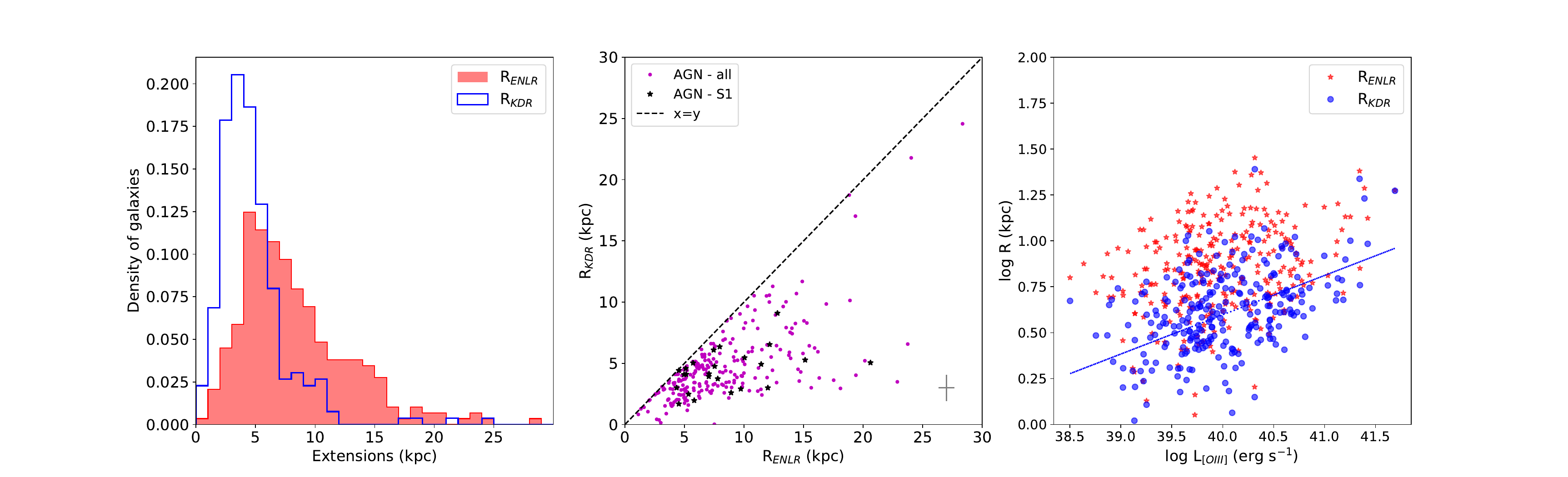}
    \caption{\textit{Left panel:} Distribution of R$_{KDR}$ (in empty blue histogram) and R$_{ENLR}$ (in filled red histogram) extents for the 274 AGN that we can calculate this parameters. \textit{Central panel:} The relation between the values of R$_{KDR}$ and R$_{ENLR}$ for each galaxy; AGN are represented by the colour pink, the black star are the S1 sample (25 AGN) and the dashed line is the equality x=y. \textit{Right panel:} Both extents vs. L[O{\sc III}]; colours follow those in the left panel, and the lines are the corresponding linear regressions. }
    \label{fig:all_EXTENSION}
\end{figure*}

The left panel of Fig. \ref{fig:all_EXTENSION} shows the distribution of the extents of the KDR and the ENLR regions for the 274 AGN that met the requirements to display both regions. The median extents are R$_{KDR}$=4.1 kpc and R$_{ENLR}$=7.2 kpc, showing that, typically, the extent of the KDR is about 57\% that of the ENLR region. The range covered by these two extents are: $0.01$\,kpc$<$R$_{KDR}$<$24.5$\,kpc and $0.7$\,kpc$<$R$_{ENLR}$<$31.9$\,kpc. 

The central panel of Fig. \ref{fig:all_EXTENSION} shows the relation between the values of R$_{KDR}$ and R$_{ENLR}$ for each galaxy.
We have identified in the figure the points corresponding to type 1 AGN as black dots. We note that they are distributed over the whole range of R$_{KDR}$ and R$_{ENLR}$ values, thus their ENLR are not particularly less extended than the dominant type 2 sources of our sample, which is somewhat unexpected within the framework of the Unified Model \citep{Antonucci85b}. If we are seeing the type 1 sources more pole-on than the type 2, and if the ENLR is bipolar, we would expect that the type 2 sources to be more extended, but we do not see this in our data, except for the most extended sources (beyond 20\,kpc) that are all type 2.

In the right panel of Fig.\ref{fig:all_EXTENSION} we show the relation between these two extents and $L_{\rm [O\:III]}$, showing that both of them increase with $L_{\rm [O\:III]}$. A linear regression fit to these relations gives:

\begin{equation}
    log(R_{KDR}) = 0.21 log(L_{\rm [O\:III]}) - 7.95
\end{equation}
\begin{equation}
    log(R_{ENLR}) = 0.12 log(L_{\rm [O\:III]}) - 4.24
    \label{eq:r_agnVSL}
\end{equation}

with correlation coefficients of $\rho = 0.4$ and $\rho = 0.2$, respectively. These correlations, although weak, at least confirm the results of Fig. \ref{fig:lX80}, supporting an increased in the extent of the kinematic disturbed region with the luminosity of the AGN.

It is interesting to compare the results above with previous ones also based on MaNGA data. \cite{Alice_paperV} find KDR and NLR extents between 0.2 -- 2.3 kpc and 0.4 -- 10.1 kpc, respectively, values lower than ours, this may be due to the difference in assumptions adopted for the KDR and NLR regions. They considered the NLR extension as the radius furthest from the nucleus within which the ratios of the [O{\sc iii}]/H$\beta$ and [N{\sc ii}]/H$\alpha$ emission lines fall in the AGN region in the BPT diagram, which leads to a smaller region than using our method when we obtained the extent of the ENLR. Regarding the distance from KDR, we attribute the difference in the values of our work and the \cite{Alice_paperV} due the fact that they used the residual velocity dispersion as a signature of kinematic disturbance, which is much less sensitive than W$_{80}$ to outflows that mostly contribute to the profile wings. \cite{Jana_paperIV} find values for the extent of the NLR in the range 1.2 -- 27 kpc using the same methodology as \cite{Alice_paperV} and using a sub-sample of 62 AGN from our sample of 293 AGN, as well as a positive correlation between R$_{ENLR}$ and $L_{\rm [O\:III]}$, in agreement with our results.
In another study, \citet{Dominika}, via the analysis of W$_{80}$ radial profiles of AGN, concluded also that higher luminosity AGN drive larger scale outflows, in agreement with our study. The relation between the ENLR extent and $L_{\rm [O\:III]}$ has also been investigated in other studies \citep{fischer_18,Thaisa_18}, that find a similar regression to that of Equation \ref{eq:r_agnVSL}.

\subsection{Kinetic impact from the AGN on the host galaxies}

In this section we discuss the kinetic impact of the AGN on the host galaxies, estimating the kinetic energy deposited on the galaxy using two methods: the first assuming that this impact is due to an AGN outflow combined with turbulence traced by the broad component; the second, using the W$_{80}$ to trace a broader feedback -- from outflows, radiation pressure and heating by the AGN radiation -- to larger scales into the galaxy.


\subsubsection{Kinetic impact calculated from the broad component}

In previous studies, the broad component of [O{\sc iii}] -- that has line widths corresponding to velocities higher than those typical of orbital motion in the galaxy -- has been frequently attributed to AGN outflows \citep[e.g.][and references therein]{Bruno_21}. We adopt this hypothesis and calculate the corresponding mass-outflow rate and kinetic power for the 135 AGN host galaxies that present the broad component in the [O{\sc iii}] emission-line profiles.

We estimate the power of the outflow to gauge its impact on the host galaxies, adopting a spherical geometry, as the data does not not have high enough angular resolution to map the geometry of the flow. We now calculate the relevant parameters as follows.

\medskip
\noindent {\bf Gas mass --} First we estimate the mass of ionised gas in the region where the broad component is observed, using the corresponding $L_{\rm [O\:III]}$ and the expression below \citep[see][]{Carniani_15}: 
\begin{equation}
    M{\rm [O\:III]}_{gas} = 0.8 \times 10^8 M_{\odot} \frac{1}{10^{[O/H]-[O/H]_\odot}} \frac{L_{\rm [O\:III]}}{10^{44} erg s^{-1}} \frac{\langle n_e \rangle}{500 cm^{-3}}, 
    \label{eq_mass}
\end{equation}
where $L_{\rm [O\:III]}$ is the integrated luminosity of the broad component in the [O\,{\sc iii}]$\lambda$5007\AA\, emission line in the region where the broad component is observed, and n$_e$ is the electron density.

In order to obtain the electron density n$_e$, we have used the ratio between the lines [S\,{\sc ii}]  $\lambda\lambda$6718,31, fitted by a single Gaussian to each line, following the expression given by \cite{Sandres_16,Kakkad_18}:
\begin{equation}
     n_e(R) = \frac{cR-ab}{a-R},
     \label{eq_n_e}
\end{equation}
where $R$ is the flux ratio $R=F([S_{II}] 6716)/ F([S_{II}] 6731)$, and a=0.4315, b=2107 and c=621.1. We compute a median value of n$_e$ per AGN. We find a mean value for the electron density of n$_e=214\pm 144.2$ \,cm$^{-3}$.

Assuming the metallicity of the gas as solar, the term of the Equation \ref{eq_mass} $[O/H]-[O/H]_\odot$ is zero \citep[eg.][]{Kakkad_superII}. We define the total mass of the gas that was disturbed by the AGN as:
\begin{equation}
     M_{gas,b} = \sum_i M{\rm [O\:III]}_{gas}^i,
     \label{eq_M_out}
\end{equation}
where the sum is performed over all spaxels showing the broad component. 

\medskip
\noindent {\bf Flow velocity --} In calculating the velocity of the gas in the region impacted by the AGN we will consider both the centroid velocity of the broad component and its velocity dispersion. We thus hereafter refer to this velocity not as the outflow velocity but as the “flow velocity”, that incorporates both outflows and other sources of disturbance in the gas, e.g. turbulence, flows perpendicular to the ionisation or outflow axis.

    \begin{equation}
         v_{flow,b} = \frac{\langle (|v_{broad}|+ 2 \sigma_{broad})\, F_{broad}\rangle}{\langle F_{broad}\rangle},
    \end{equation}
where $v_{broad}$ is its centroid velocity (velocity shift relative to the galaxy systemic velocity \cite{Bruno_21}), $\sigma_{broad}$ is the velocity dispersion of the broad component (thus taking also into account the impact from the gas turbulence), and F$_{broad}$ is the integrated flux of the [O\,{\sc iii}] broad component over the region it is observed. This equation is a ``weighted" average of the velocity over the region of the kinetic impact probed by the broad component.

\medskip
\noindent {\bf Radius of kinetic impact --} Another necessary parameter in the calculation is the radius of the impact region. We adhere to the methodology outlined by \cite{rogemar_23}, where this radius is the weighted average of the distances from the nucleus of the spaxels presenting the broad component:
\begin{equation}
     R_{b} = \frac{\langle R_{broad} F_{broad}\rangle}{\langle F_{broad}\rangle},
     \label{eq:r_out_b}
\end{equation}
Here, $R_{b}$ represents the resulting radius of the region showing the broad component, $R_{broad}$ denotes the distances of individual spaxels from the nucleus and $F_{broad}$ represents the corresponding fluxes in the broad component. The angular brackets denote averaging over all spaxels.

\medskip
\noindent{\bf Mass flow rate and power --} The mass flow rate ($ \dot{M}_{flow}$) is computed assuming a spherical geometry, and is given by:
\begin{equation}
    \dot{M}_{flow,b} = \frac{M_{gas} v_{flow,b}}{ R_{b}}
    \label{eq:dotM}
\end{equation}
This expression has been used recently in several works \citep{Kakkad_superII,Bruno_21,kakkad_2022,rogemar_23}, and allows us to estimate the mass-flow rate of an approximately isotropic flow.
The kinetic power of the flow ($ \dot{E}_{out}$) is calculated as:
\begin{equation}
    \dot{E}_{flow,b} = \frac{1}{2} \dot{M}_{flow,b} v_{flow,b}^2
    \label{eq:dotE}
\end{equation}


\subsubsection {Kinetic impact calculated from W$_{80}$}

The W$_{80}$ parameter, being the width of the profile corresponding to 80\% of the line flux, is sensitive to high velocity gas contributing to the profile wings, including the cases with broad components discussed above. But W$_{80}$ is also sensitive to increased velocity dispersion. Thus W$_{80}$ can be an indicator of gas outflows but can also trace the presence of turbulence in the gas, e.g. via increased velocity dispersion that can be isotropic, along or perpendicular to the ionisation axis or radio jet \citep{rogemar_23, Dutra_AGNIFS,Venturi_2021,2021_Rifel_chemicalAbundance}. As there is a clear enhancement of W$_{80}$ in AGN relative to the control galaxies, the increased turbulence is also due to the AGN, considering also the fact that in section \ref{bojo} we showed that this increase is not due to the gravitational potential of the galaxy bulge. Therefore, W$_{80}$ is a tracer of the kinetic impact of the AGN on the ambient gas, and we now use its value for the hole AGN sample to compare the resulting properties of the flows with those obtained using the broad component.

We maintain the assumption of a spherical geometry, but now considering the presence of a kinematic disturbance over the entire KDR – defined as the area where $W_{80}\ge W_{80,cut}$ = 315 km\,s$^{-1}$ and the line ratios trace AGN excitation (Section \ref{kdr}).

\medskip
\noindent {\bf Gas mass --} We use equation \ref{eq_mass} to calculate the mass in the flow from $L_{\rm [O\:III]}$. However, instead of using the total flux of the [O{\sc III}] emission line to calculate its luminosity, we adopt and conservative approach integrating only the part where the absolute velocities are larger than $W_{80,cut}/2$ relative to the peak velocity, at each spaxel of the KDR. We then add this contribution from all spaxels of the KDR \citep{rogemar_23,Dutra_AGNIFS}. This assumption excludes the contribution of the lower velocity gas, that should be dominated by motions in the galaxy potential.

\medskip
\noindent{\bf Flow velocity from W$_{80}$--} Previous works have been using W$_{80}$ as an kinetic feedback tracer in different ways \citep{rogemar_22,Bruno_21,Dutra_AGNIFS,Dominika}, here we follow \cite{rogemar_23}, adopting as the flow velocity:
    \begin{equation}
v_{flow,W_{80}} = \frac{\langle W_{80,KDR} F_{KDR}\rangle}{\langle F_{KDR}\rangle},
    \end{equation}
\noindent which is a ``weighted" average of $W_{80}$ over the KDR spaxels \citep{rogemar_23}, where $\langle F_{KDR} \rangle$  is the average flux of the KDR spaxels.

\medskip
\noindent{\bf Radius of kinetic impact--} in order to determine this radius, we modify Equation \ref{eq:r_out_b} to obtain a weighted average of the radius of the KDR, derived as:

\begin{equation}
     R_{flow,W_{80}} = \frac{\langle R_{KDR} F_{KDR}\rangle}{\langle F_{KDR}\rangle}.
\end{equation}

Finally, we adhere to the method described earlier to determine the mass flow rate $\dot{M}_{flow,W80}$ and power $\dot{E}_{flow,W80}$, using modifications of eqs.\,\ref{eq:dotM} and \ref{eq:dotE} to take into account the contribution from all the spaxels of the KDR.

\subsubsection{Comparing the two methods}
\begin{figure*}
    \centering
    \includegraphics[trim = 3mm 0mm 3mm 0mm,clip,width=1\textwidth]{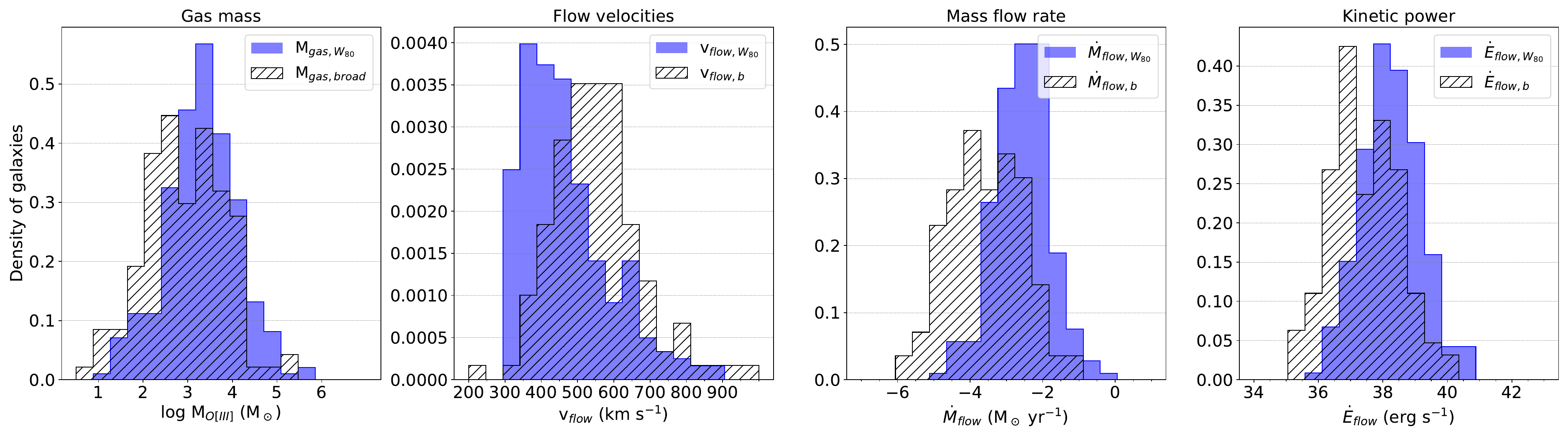}
    \caption{Histograms comparing the flow proprieties for 135 AGN hosts (that present broad components)  
    via two methods: using the broad component     (hatched), and using the W$_{80}$ as measure of the velocity of the flow (in blue). From left to right: (1) the [O\,{\sc iii}] gas mass in flow;     (2) the flow velocities (v$_{flow,b}$ and $v_{flow,W_{80}}$); (3) mass flow rates ($\dot{M}_{flow,b}$ and $\dot{M}_{flow,W_{80}}$); (4) kinetic power of the flow ($\dot{E}_{flow,b}$ and $\dot{E}_{flow,W_{80}}$). The p-values of the k-s test for the distributions are $10^{-5}$,$10^{-8}$, $10^{-13}$ and $10^{-9}$, respectively.}
    \label{fig:propri_out}
\end{figure*}

The key findings regarding the properties of the AGN influence in the host galaxy for the two methods described above are illustrated in Figure \ref{fig:propri_out}, obtained for 135 AGN hosts (those presenting the broad component). From left to right we show histograms 
for the gas mass, flow velocity, mass flow rate, and kinetic power of the flow. Results obtained using the W$_{80}$ method are represented in blue, while those obtained using the broad component method are represented in hatched black.
   
In the first panel of Figure~\ref{fig:propri_out}, we present the results for the ionised gas masses from Equation \ref{eq_M_out} using the two methods. The resulting masses are in the range: $10^{0.5}\le M_{gas,b} \le-10^{5.5},M_\odot$, and $10^{1}\le M_{gas,W_{80}}\le 10^{6} M_\odot$, with mean log(M/M$\odot$) values of $3.7 \pm 4.2$ and $4.3 \pm 5$, respectively. The $k-s$ test for this distribution yields a p-value of $\approx 10^{-5}$, confirming that they are different, as can readily be seen. This can be attributed to the lower amplitude of the broad component and to the fact that the region presenting the broad component is much less extended than the KDR.

In the second panel of figure \ref{fig:propri_out}, showing the flow velocity distributions, the range of values are similar, although with usually higher values derived via the broad component method than via the W$_{80}$ method. The mean values are v$_{flow,b} = 593.7 \pm 204.5$ km\,s$^{-1}$ and v$_{flow,W_{80}} = 472.9 \pm 135.5$ km\,s$^{-1}$.
The distribution of mass flow rates in the third panel show higher values for the W$_{80}$ method, with means of  $\dot{M}_{flow,b}=3.2 \times 10^{-3}$\,M$_\odot$\,yr$^{-1}$ and $\dot{M}_{flow,W_{80}}1.3\times$10$^{-2}$\,$M_\odot$\,yr$^{-1}$.

Finally, for the kinetic power of the flow we find mean values of $\dot{E}_{flow,b}=4.2 \times 10^{38}$\,erg\,s$^{-1}$ and $\dot{E}_{flow,W_{80}}=1.5 \times 10^{39}$\,erg\,s$^{-1}$. 

The differences in the mass flow rates and powers are quantified by the k-s test with p-values of $\sim 10^{-9}$ and  $10^{-13}$, respectively. Despite the mostly higher values of the flow velocity for the broad component, the corresponding ionised gas masses are smaller than that obtained via the W${80}$ method, resulting in lower values of $\dot{M}{out,b}$ and $\dot{E}_{out,b}$. 

Uncertainties in the above quantities include the adopted geometry for the region of the galaxy impacted by the AGN (that can be, e.g. bi-conical, spherical), although recent studies have shown that the choice of geometry result in overall similar values of the mass-flow rates \citep{kakkad_2022,rogemar_23}. 

Another uncertainty is in the [O\,{\sc iii}] emission-line fluxes used to obtain the perturbed gas masses. In the broad component method we use only the flux of the broad component, considered to correspond to the perturbed gas, an assumption frequently adopted in the literature  
 \citep[eg.][]{Bruno_21,kakkad_2022,rogemar_23}. 
 
In the W$_{80}$ method, we assume that the perturbed gas corresponds to velocities along the line profiles that are above W$_{80,cut}$/2). This parameter and the extent of the KDR evidences that the region perturbed by the AGN is much more extended than the region probed by the broad component.

\begin{figure}
    \centering
    \includegraphics[width=\columnwidth]{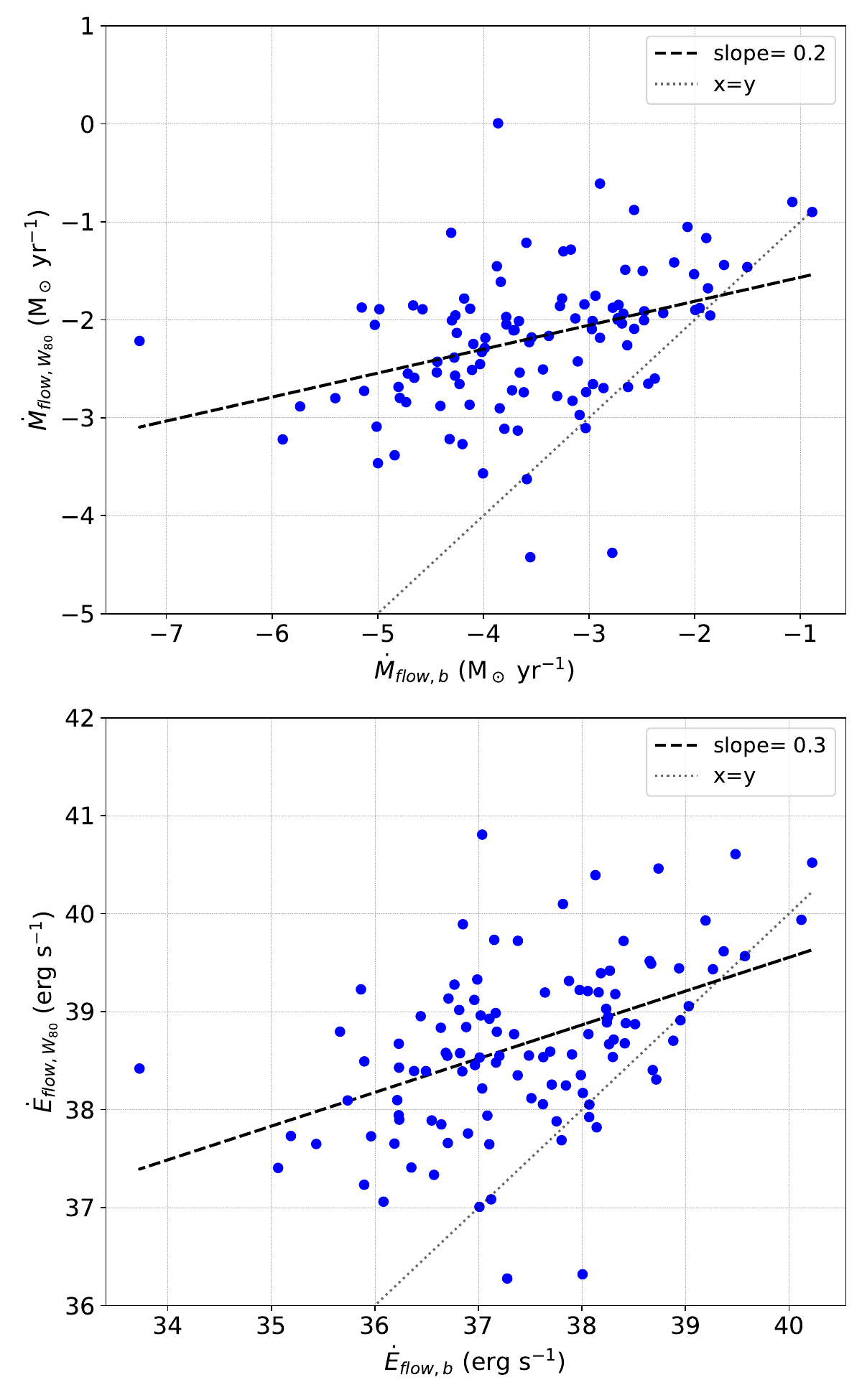}
    \caption{Comparison between the values of the mass outflow rate (upper panel) and power (bottom panel) obtained using the two methods -- broad component and W$_{80}$. The dashed line represent the linear regression and the dotted line represents x=y.}
    \label{fig:broad_w80}
\end{figure}

The comparison between $\dot{M}$ and $\dot{E}$ obtained via the two methods is shown in Figure \ref{fig:broad_w80} for the 135 AGN that have two components. The dotted line near the bottom shows the unitary relation, evidencing that the values obtained via the W$_{80}$ method are mostly higher than those obtained via the broad component method, with the ratios between the two being very low for low values and increasing as the values increase: the ratios $\dot{M_b}/\dot{M_{W80}}$ and $\dot{E_b}/\dot{E_{W80}}$ range from 10$^{-3}$ for the lowest $\dot{M_b}$ and $\dot{E}$ to almost 1 for the highest values. A linear regression to the points has a slope of 0.2 and a Spearman's rank correlation coefficient of 0.32 in both cases, showing a weak correlation.

The above results can be explained as follows:  while the broad component is probably tracing an actual outflow from the AGN, W$_{80}$ traces a broader and more extended disturbance, not only due to a direct outflow but also from turbulence in the gas due to lateral disturbance produced by an outflow or jet passing through the ISM or to ISM heating and pressure produced by the AGN radiation. 

We thus use the W$_{80}$ method to calculate $\dot{M}$ and $\dot{E}$ for the whole AGN sample as discussed in the next section. 

\subsection{$\dot{M}_{flow}$ and $\dot{E}_{flow}$ vs. L$_{bol}$}

We now investigate the relation of AGN values of $\dot{M}_{flow}$ and $\dot{E}_{flow}$ as a function of L$_{bol}$ (obtained as described in section \ref{sec:L_bol}) and compare it with those from the literature \citep[e.g.][]{Fiore}. 

\begin{figure*}
    \centering
    \includegraphics[trim = 25mm 5mm 25mm 20mm,clip,width=1\textwidth]{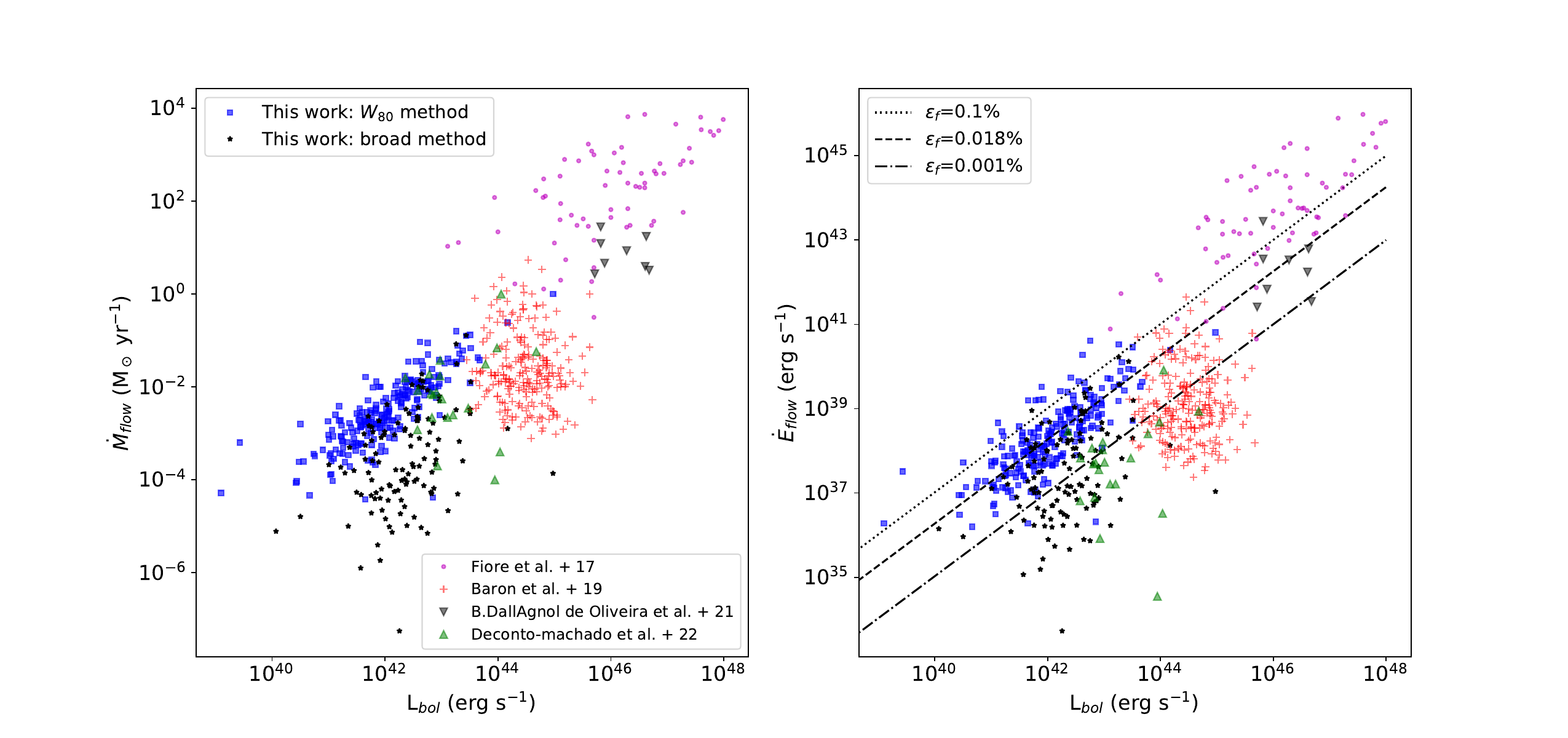}
    \caption{Mass flow rate $\dot{M}_{flow}$ and kinetic power $\dot{E}_{flow}$ as a function of the AGN bolometric luminosity L$_{bol}$. \textit{Left panel:} Results for mass flow rate where $\dot{M}_{flow,b}$ are represented by black stars and $\dot{M}_{flow,b}$ as dark blue squares. The data from literature as identified in the inserted bottom rectangle;
    \textit{Left panel:} Kinetic power, where the symbols and colours are the same as in left panel. The dotted, dashed and dot-dashed black lines correspond to kinetic coupling efficiencies $\epsilon_f$ of 0.1\%, 0.017\% and 0.001\%, respectively.}
    \label{fig:M_E_out_VS_L_bol}
\end{figure*}

Figure \ref{fig:M_E_out_VS_L_bol} displays this relation, where the values obtained from the method of the broad component are identified as black stars and those from the W${80}$ method are shown as blue squares. While the values using the two components are only available for 135 AGN, the values derived from W${80}$ were obtained for the whole AGN sample. Included in figure \ref{fig:M_E_out_VS_L_bol} are also results from \cite{Bruno_21},\cite{Fiore}, \cite{baron} and \cite{Alice_paperV}, as comparison. 

We confirm the positive correlation between $\dot{M}_{flow}$ and $\dot{E}_{flow}$ with L$_{bol}$ found in previous studies \citep[e.g.][]{Fiore,Dominika}. Figure \ref{fig:M_E_out_VS_L_bol} shows also that our MaNGA AGN approximately follow the relation between $\dot{M}_{flow}$, $\dot{E}_{flow}$ and L$_{Bol}$ observed in these previous studies of more luminous sources, occupying the region corresponding to the low-luminosity end of the relation.

The values obtained using the broad component spread to lower values of $\dot{M}$ and $\dot{E}$. As discussed in the previous section, we believe the W$_{80}$ method is a better tracer of AGN feedback than the broad component method, at least for the low-luminosity AGN that dominate our MaNGA sample. This is because it traces not only the effect of AGN outflows, but also of turbulence caused by the lateral expansion of such outflows as well as radiation heating and pressure on the galaxy ISM.

To evaluate the impact of these AGN flows on the host galaxies, we calculated the coupling efficiency -- $\epsilon_f=\dot{E}/L_{bol}$ for the 293 AGN, as shown in Figure \ref{fig:M_E_out_VS_L_bol}. The diagonal lines correspond to kinetic coupling efficiencies $\epsilon_f$ of 0.1\% (dotted line), 0.018\% (dashed line) and 0.001\% (dot-dashed line), the two latter corresponding to the median values of our AGN sample using the W$_{80}$ and broad component methods, respectively.
For the $W_{80}$ method, 158 AGN reach values of $\epsilon_f$ above 0.01\%.

These low coupling efficiencies are apparently not powerful enough to suppress star formation in the host galaxies, as models such as those of \citet{di_matteo} and \citet{hopkins_10} require a minimum $\epsilon_f$ in the range 0.5\% to 5\% for this to happen. 

Nevertheless, more recent studies \citep[e.g.][]{Harrison_18,Ward_22} argue that most AGN are found in star-forming galaxies, and that low $\epsilon_f$ values do not preclude a kinematic feedback effect on the host galaxies. Even if such mild feedback does not immediately halt star formation, the continuous heating and disturbance of the ambient gas produced by the AGN will result in a cumulative feedback effect on the host galaxy \citep{Piotrowska_22}, in what may be called a ``maintenance mode" feedback, that will influence its star formation pace in the long range.

In addition, what we have calculated here is the kinetic feedback power of just the ionised gas phase. Other gas phases, such as the molecular gas, may add to this power. As pointed out in studies such as that of \citet[][and references therein]{Harrison_18,rogemar_23} the kinetic feedback produced by AGN flows is just about 20\% of the AGN feedback power on the host galaxy \citep{Richings_18}. 
The relation we have found between W$_{80}$ and $L_{\rm [O\:III]}$ (Figure\,\ref{fig:lX80}) extending down to the lowest luminosities, and the extents of the KDR that reach on average half the extent of the ENLR (which many times extend out to the borders of the galaxy) show that even low-power AGN, such as those of the present study, can have a significant effect on the host galaxy, disturbing its gas out to several kiloparsecs into the galaxy.


\section{Summary and Conclusions}
\label{sec:6}

In this work, we have obtained the ionised gas kinematics measured via the [O\,{\sc iii}]$\lambda5007$ emission-line profile of a sample of 293 active galaxies as compared to those of a control sample of 485 non-active galaxies from the MaNGA survey. Each AGN host has two control galaxies matched in morphology, redshift, inclination, total stellar mass and stellar velocity dispersion. We have fitted the line profiles using one or two Gaussian curves with the main goal to obtain the non parametric value of W$_{80}$ as a tracer of AGN kinematic disturbances and compare its value between AGN and controls. We define a Kinematically Disturbed Region -- KDR, corresponding to the region with W$_{80}$ higher than those typical of the control galaxies and compare its extent with that of the ENLR region.

We compute the kinetic feedback of the AGN on the host galaxies in two ways: one considering only the broad component -- for the 153 AGN showing two components, and the other using the W$_{80}$ values for the hole sample, under the assumption that the KDR traces the kinetic feedback of the AGN on the ambient gas.

The main conclusions of this work are:

\begin{itemize}
    \item The W$_{80}$ [O\,{\sc iii}] values in the inner few kpc are usually larger in the AGN hosts than in the controls; we attribute this difference as due to a kinematic disturbance due to the AGN;
    
    \item The weighted mean $\langle W_{80}\rangle$ values for the AGN range from 200\,km\,s$^{-1}$ to 900\,km\,s$^{-1}$, with a median value of 453\,km\,s$^{-1}$; for the control galaxies, the range is 150--600\,km\,s$^{-1}$, with a median value of 306\,km\,s$^{-1}$; the median difference between AGN and controls is 238\,km\,s$^{-1}$;
    
    \item We find a positive correlation between $\langle W_{80}\rangle$ and  $L_{\rm [O\:III]}$ for the AGN that, combined with the control sample gets improved. This could be due to a possible contribution from a faint AGN in at least part of the control galaxies;

    \item We have ruled out the possibility that the higher W$_{80}$ values of the AGN relative to controls are due to more massive bulges in the AGN, as we show that the bulge velocity dispersions are similar for the two samples as well as their photometric properties;
    
    \item The KDR extent R$_{KDR}$ ranges from 0.01\,kpc to 24.5\,kpc, while the ENLR extent R$_{ENLR}$ ranges from 0.7\,kpc to 31.9\,kpc kpc; we obtain a positive correlation between these two extents, with a mean ratio $\langle R_{KDR}/R_{ENLR}\rangle = 57\%$; both extents correlate with  $L_{\rm [O\:III]}$;

    \item We did not find any difference in terms of AGN luminosity, W$_{80}$ and extents R$_{ENLR}$ and R$_{KDR}$ for the 25 type 1 AGN in our sample relative to the dominant type 2 sources. These results support that the kinematic disturbances we are probing are mostly isotropic, at least in the scales we are probing.  We thus estimate its effect on the galaxy as due to an approximately isotropic gas flow;
    
    \item We have estimated the mass ﬂow rate $\dot{M}_{flow}$ and kinetic power $\dot{E}_{flow}$ for the sub-sample of 135 AGN that showed a broad component in two different methods: one using the broad component and the other using the W$_{80}$ values; the resulting mass flow rates are in the range $10^{-4} \le \dot{M}_{flow,W_{80}} \le 1$ \,M$_\odot$\,yr$^{-1}$ and can reach 1 to 2 orders of magnitude lower for $\dot{M}_{flow,b}$;

    \item We conclude that W$_{80}$ is a better tracer of the AGN feedback than the broad component, that is observed only in the innermost region, while broadening of W$_{80}$ is observed out to several kpc and probe not only outflows, but other sources of kinetic impact, such as turbulence due to lateral expansion of the gas pushed by outflows, radiation heating and pressure;  

    \item Using the W$_{80}$ method, the kinetic powers of the AGN have a mean coupling efficiency of 0.02\% L$_{bol}$ and fit into the low-luminosity end of the proportionality relation between log($\dot{M}$) and log($\dot{E}$) with log($L_{bol}$), known from the literature for more luminous sources.
  
\end{itemize}

In spite of the low kinetic power of the AGN in our MaNGA sample, the relation between $\langle W_{80}\rangle$ and L[O{\sc iii}], together with the large extents of the KDR, indicate that these low-luminosity AGN do impact the host galaxy. The combination of low power and large distances into the galaxy being impacted by the AGN configure what can be described as a ``maintenance mode feedback" occurring in these AGN.

\section*{Acknowledgements}

This study was funded by the Coordenação de Aperfeiçoamento de Pessoal de Nível Superior - Brasil (CAPES) - Finance Code 001.

TSB and RAR acknowledges financial support from Conselho Nacional de Desenvolvimento Cient\'ifico e Tecnol\'ogico (Projs. 425966/2016-0 and 303450/2022-3) and Funda\c c\~ao de Amparo \`a pesquisa do Estado do Rio Grande do Sul (Projs.  16/2551-0000495-1 and 21/2551-0002018-0).

RR acknowledges support from the Fundaci\'on Jes\'us Serra and the Instituto de Astrof{\'{i}}sica de Canarias under the Visiting Researcher Programme 2023-2025 agreed between both institutions. RR, also acknowledges support from the ACIISI, Consejer{\'{i}}a de Econom{\'{i}}a, Conocimiento y Empleo del Gobierno de Canarias and the European Regional Development Fund (ERDF) under grant with reference ProID2021010079, and the support through the RAVET project by the grant PID2019-107427GB-C32 from the Spanish Ministry of Science, Innovation and Universities MCIU. This work has also been supported through the IAC project TRACES, which is partially supported through the state budget and the regional budget of the Consejer{\'{i}}a de Econom{\'{i}}a, Industria, Comercio y Conocimiento of the Canary Islands Autonomous Community. RR also thanks to Conselho Nacional de Desenvolvimento Cient\'{i}fico e Tecnol\'ogico  ( CNPq, Proj. 311223/2020-6,  304927/2017-1 and  400352/2016-8), Funda\c{c}\~ao de amparo \`{a} pesquisa do Rio Grande do Sul (FAPERGS, Proj. 16/2551-0000251-7 and 19/1750-2), Coordena\c{c}\~ao de Aperfei\c{c}oamento de Pessoal de N\'{i}vel Superior (CAPES, Proj. 0001).

G.S.I. acknowledges financial support from FAPESP (Fundação de Amparo à Pesquisa do Estado de São Paulo, Proj. 2022/11799-9).

\section*{Data Availability}
The data underlying this article are available under SDSS collaboration rules.


\bibliographystyle{mnras}
\bibliography{example} 







\bsp	
\label{lastpage}
\end{document}